\documentclass[a4paper,11pt]{article}
\pdfoutput=1 
\usepackage{jheppub} 
\usepackage[T1]{fontenc} 
\usepackage{amsmath,amssymb,epsfig,ulem,color,slashed,wasysym,comment,txfonts,relsize}
\allowdisplaybreaks 

\def \beq{\begin{equation}}
\def \eeq{\end{equation}}
\def \bea{\begin{eqnarray}}
\def \eea{\end{eqnarray}}

\title{Let there be light from a second light Higgs doublet}

\author[1,2]{Ulrich Haisch}

\author[1]{and Augustinas Malinauskas}

\affiliation[1]{Rudolf Peierls Centre for Theoretical Physics,
   University of Oxford, OX1 3NP Oxford, United Kingdom}
    
\affiliation[2]{CERN, Theoretical Physics Department,  CH-1211 Geneva 23, Switzerland}

\emailAdd{ulrich.haisch@physics.ox.ac.uk}
\emailAdd{augustinas.malinauskas@physics.ox.ac.uk}

\abstract{

\noindent In this article, we demonstrate that the unexpected peak at around $95 \, {\rm GeV}$ as seen recently by  CMS in the di-photon final state can be explained within the type-I~two-Higgs-doublet model by  means of a  moderately-to-strongly fermiophobic CP-even Higgs $H$. Depending on the Higgs mass spectrum, the production of such~a~$H$ arises dominantly from vector boson fusion or through a cascade in either $pp \to t \bar t$ with $\overset{\text{\relsize{-2}(}-\text{\relsize{-2})}}{t} \to H^{\pm} \overset{\text{\relsize{-2}(}-\text{\relsize{-2})}}{b} \to W^{\pm \, \ast} H \overset{\text{\relsize{-2}(}-\text{\relsize{-2})}}{b}$ or $pp \to A$ with $A \to W^{\mp} H^{\pm} \to W^{\mp} W^{\pm} H$ or via $pp \to W^{\pm \, \ast} \to H^\pm H$.  In this context, we  also discuss other Higgs anomalies such as the LEP excess in Higgsstrahlung and the observation of enhanced  rates  in $t \bar t h$ at both the Tevatron and the LHC, showing that parameters capable of explaining the CMS di-photon signal can address the latter deviations as well. The Higgs spectra that we explore comprise masses between $80 \, {\rm GeV}$ and $350\, {\rm GeV}$. While at present all constraints from direct and indirect searches for spin-0 resonances can be shown to be satisfied for such light Higgses, future LHC data will be able to probe the parameter space that leads to a simultaneous explanation of the discussed anomalies.}

\preprint{CERN-TH-2017-276}

\begin{document} 

\maketitle

\flushbottom

\section{Introduction}
\label{sec:introduction}

The search for the standard model (SM) Higgs has a long history. It started at LEP, continued at the Tevatron and culminated in 2012 with the discovery of a spin-0 resonance $h$ with a mass of around $125 \, {\rm GeV}$ at the LHC. In the last five years the LHC Higgs program has matured~\cite{Khachatryan:2016vau}, providing precise measurements of processes   such as $pp \to h \to \gamma \gamma$ and $pp \to h \to ZZ^\ast \to 4 \ell$ (for the latest LHC results at $\sqrt{s} = 13 \, {\rm TeV}$ see~\cite{ATLAS-CONF-2017-045,ATLAS-CONF-2017-032,CMS-PAS-HIG-16-040,Sirunyan:2017exp}) with SM rates of order $0.1 \, {\rm pb}$ and below. 

Although the  $125 \, {\rm GeV}$ spin-0 resonance has properties very close to the one expected for the SM Higgs, it is still well possible that a non-minimal Higgs sector is realised in nature while $h$ is  SM-like. Searches for additional Higgs-like particles have been performed at all major high-energy colliders and more than once deviations from the SM predictions or signs of new resonances were found. While some of these excesses --- such as the infamous $750 \, {\rm GeV}$ peak in the di-photon mass spectrum  reported in 2016 by both ATLAS and CMS~\cite{ATLAS-CONF-2016-059,CMS-PAS-EXO-16-027} --- disappeared with the collection of more data, other anomalies remained and new ones emerged. Examples of lasting anomalies  are the LEP excess in Higgsstrahlung~\cite{Barate:2003sz}, the measurement of enhanced rates in $t \bar t$ associated Higgs production at  the Tevatron~\cite{Collaboration:2012bk,Aaltonen:2013ipa} as well as at the~LHC~\cite{ATLAS-CONF-2016-058,CMS-PAS-HIG-16-022}.  In all three cases the significance of the observed deviation is  at the level of $2\sigma$. Most recently, the observation of an unexpected bump at low mass in the di-photon final state was reported by the CMS collaboration~\cite{CMS-PAS-HIG-17-013},  increasing the significance of earlier $8 \, {\rm TeV}$ results~\cite{CMS-PAS-HIG-14-037}.  The combined local (global) significance of the CMS di-photon excess is $2.8\sigma$ ($1.3\sigma$). 

While none of the aforementioned deviations is by itself statistically significant, it seems worthwhile to ask whether these anomalies might be related within a specific extension of the~SM. In this article, we consider the type-I two-Higgs-doublet model~(2HDM) and show that this model can provide a simultaneous explanation of several of the observed excesses in terms of a moderately-to-strongly fermiophobic CP-even Higgs $H$ with a mass of about $95 \, {\rm GeV}$. The fermiophobic nature of the resonance leads to  unconventional production mechanisms, including $H$ production through cascade decays of charged Higgses $H^\pm$ or neutral CP-odd states $A$ and associated $H^\pm H$ production, with $gg \to H$  always accounting only for  a subleading part of the total rate. Also the decays of the~$H$ and the other spin-0 states turn out to have unfamiliar features, which we illustrate by discussing four different benchmark scenarios. These benchmark scenarios all have in common that they feature a light spectrum of Higgses with masses not exceeding $350 \, {\rm GeV}$. We explicitly show that in all four cases the chosen parameters are compatible with the existing direct and indirect constraints on the type-I~2HDM parameter space.  While this study was ongoing, a~similar investigation of the CMS di-photon excess in the context of the type-I 2HDM has been presented in~\cite{Fox:2017uwr}. Whenever indicated we will highlight the similarities and differences between this and our work.

The outline of this article is as follows. In Section~\ref{sec:typeI} we first recall briefly the structure of the type-I~2HDM and then discuss in Section~\ref{sec:analysis} four benchmark scenarios that render consistent explanations of the recent CMS di-photon excess as well as some of the LEP, Tevatron and the other LHC anomalies mentioned above. For each benchmark scenario we also discuss strategies of how-to test it at future LHC runs. Our conclusions are presented in~Section~\ref{sec:conclusions}.

\section{Type-I~2HDM in a nutshell}
\label{sec:typeI}

The 2HDM scalar potential that we will consider throughout this work is given by the following expression (see for example~\cite{Gunion:1989we,Branco:2011iw} for a review)
\beq \label{eq:VH}
\begin{split}
V_H & = \mu_1 H_1^\dagger H_1 + \mu_2 H_2^\dagger H_2 + \left ( \mu_3  H_1^\dagger H_2 + {\rm h.c.} \right ) + \lambda_1  \hspace{0.25mm} \big ( H_1^\dagger H_1  \big )^2  + \lambda_2  \hspace{0.25mm} \big ( H_2^\dagger H_2 \big  )^2 \\[2mm]
& \phantom{xx} +  \lambda_3 \hspace{0.25mm} \big ( H_1^\dagger H_1  \big ) \big ( H_2^\dagger H_2  \big ) + \lambda_4  \hspace{0.25mm} \big ( H_1^\dagger H_2  \big ) \big ( H_2^\dagger H_1  \big ) + \left [ \lambda_5   \hspace{0.25mm} \big ( H_1^\dagger H_2 \big )^2 + {\rm h.c.} \right ]  \,.
\end{split} 
\eeq
Here we have imposed a $Z_2$ symmetry under which $H_1 \to H_1$ and $H_2 \to -H_2$. The parameters $\mu_{1,2}$ and $\lambda_{1,2,3,4}$ are real, while $\mu_3$ and $\lambda_5$ are in general complex. To avoid possible issues with electric dipole moments, we assume in what follows that $\mu_3$ and $\lambda_5$ have no imaginary parts. This automatically ensures that the potential is CP conserving,~i.e.~the mass eigenstates have definite CP properties. In addition, by appropriately charging the right-handed fermions, the $Z_2$ symmetry can also be used to obtain one of the four 2HDMs with natural flavour conservation, eliminating phenomenologically dangerous tree-level flavour-changing neutral currents.  The discrete symmetry is however softly broken by the term $\mu_3  H_1^\dagger H_2 + {\rm h.c.}$ The vacuum expectation values~(VEVs) of the Higgs doublets are given by $\langle H_i \rangle = (0,v_i/\sqrt{2})^T$ with $v = \sqrt{v_1^2 + v_2^2} \simeq 246 \, {\rm GeV}$ the electroweak VEV and we  define $\tan \beta = v_2/v_1$. 

The potential~(\ref{eq:VH}) gives rise to five physical spin-0 states: two neutral CP-even ones ($h$ and $H$), one neutral CP-odd state ($A$), and the remaining two carry electric charge of $\pm 1$ and are degenerate in mass ($H^\pm$). We identify the $125 \, {\rm GeV}$ resonance discovered at the LHC with the CP-even Higgs~$h$ while the masses of the other scalars are free parameters. The angle that mixes the neutral CP-even weak eigenstates into the mass eigenstates $h$ and $H$ will be denoted by $\alpha$. Diagonalising the mass-squared matrices of the scalars leads to relations between the fundamental parameters of $V_H$ and the physical masses and mixing angles.  This allows one to trade the parameters $\mu_1$, $\mu_2$, $\mu_3$, $\lambda_1$, $\lambda_2$, $\lambda_4$, $\lambda_5$ for $\alpha$, $\beta$, $M_h$, $M_H$, $M_A$, $M_{H^+}$ and $v$.  The only remaining free parameter of the original Higgs potential entering our calculations is $\lambda_3$. We will use it together with the latter parameters as input in our numerical analysis. 

In all 2HDMs  with CP conservation the tree-level couplings of the CP-even Higgs mass eigenstates to gauge bosons are given relative to the coupling of the SM Higgs by 
\begin{equation} \label{eq:kappaV}
\kappa^h_V = \sin \left ( \beta - \alpha \right ) \,, \qquad 
\kappa^H_V = \cos \left ( \beta - \alpha \right ) \,,
\end{equation}
where $V = W, Z$. The fermion couplings to $h$, $H$, $A$ and $H^+$ instead depend on the specific realisation of the Yukawa sector.  In the type-I~2HDM the neutral Higgs couplings are 
\begin{equation}  \label{eq:kappaftypeI}
\kappa^h_f = \frac{\cos \alpha}{\sin \beta}  \,, \qquad 
\kappa^H_f = \frac{\sin \alpha}{\sin \beta}  \,,  \qquad 
\kappa^A_u =  -\kappa^A_{d,\ell} =  \cot \beta \,,
\end{equation}
relative to the SM and the couplings of the charged Higgses to fermions resemble those of the CP-odd Higgs. Notice that the interactions of $h$ become SM-like,~i.e.~$\kappa^h_f \to 1$,  in the limit $\alpha \to 0$ and $\beta \to  \pi/2$. Furthermore, the CP-even Higgs $H$ does not couple to fermions~(i.e.~fermiophobic) for~$\alpha = 0$, while it does not couple to gauge bosons (i.e.~gaugephobic) for $\alpha = \beta \pm \pi/2$.

\begin{figure}[!t]
\begin{center}
\includegraphics[width=0.975\textwidth]{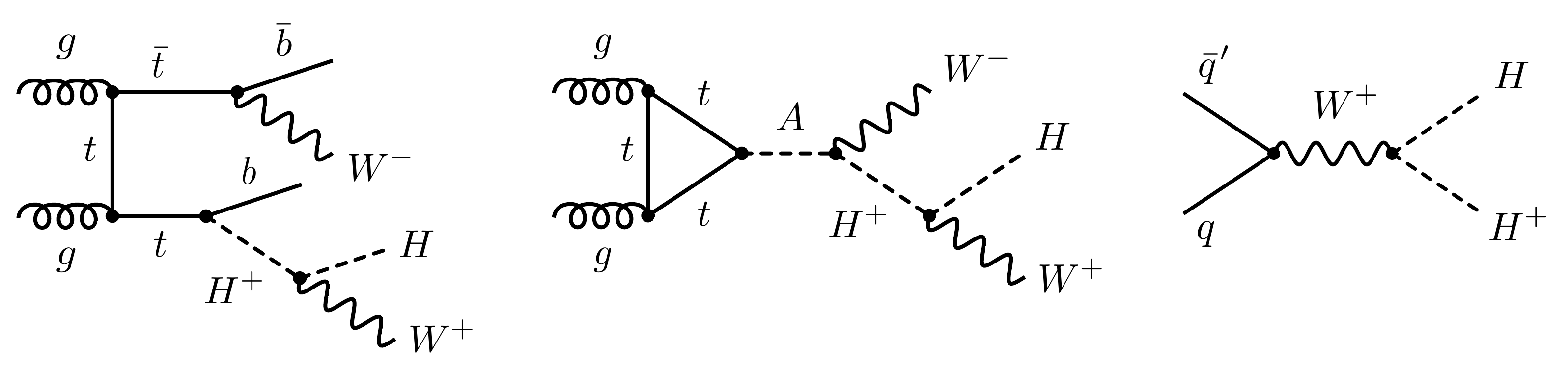}  
\vspace{2mm}
\caption{\label{fig:1}  Exotic $H$ production channels through cascades or in association with a charged Higgs. The left Feynman diagram shows the process $gg \to t \bar t$ followed by $t \to H^+ b \, (H^+ \to W^+ H)$, the middle graph illustrates the reaction $gg \to A$ with $A \to W^-  H^+ \, (H^+ \to W^+ H)$, while the diagram on the  right corresponds to the transition $q \bar q^{\hspace{0.25mm} \prime} \to W^+ \to H^+ H$.   }
\end{center}
\end{figure}

\section{Numerical analysis}
\label{sec:analysis}

In the following, we will show that the type-I~2HDM provides an economic explanation of the small CMS excess in the di-photon mass spectrum at around $95 \, {\rm GeV}$~\cite{CMS-PAS-HIG-17-013,CMS-PAS-HIG-14-037} in terms of a moderately-to-strongly fermiophobic $H$,~i.e.~models with small values of $\alpha$. We find that depending on the choice of mixing angles $\alpha$ and $\beta$ as well as the masses $M_A$ and $M_{H^+}$, the production of such a~$H$ proceeds dominantly either via the vector boson fusion (VBF) and associated (WH and ZH) channels~\cite{Cacciapaglia:2016tlr} or through a cascade in either $pp \to t \bar t$ with $\overset{\text{\relsize{-2}(}-\text{\relsize{-2})}}{t} \to H^{\pm} \overset{\text{\relsize{-2}(}-\text{\relsize{-2})}}{b} \to W^{\pm \, \ast} H \overset{\text{\relsize{-2}(}-\text{\relsize{-2})}}{b}$~\cite{Alves:2017snd} or  $pp \to A$ with $A \to W^{\mp} H^{\pm} \to W^{\mp} W^{\pm} H$. If the charged Higgs is very light associated $H$ production via $pp \to W^{\pm \, \ast} \to H^{\pm} H$~\cite{Akeroyd:2003bt,Akeroyd:2003xi} can also be the most important  production mode. Gluon~fusion~(ggH)  instead accounts only for a subleading fraction of the total $H$ production in all cases. Examples of Feynman graphs that can give rise  to $H$ production in 2HDMs via a cascade or in association with a charged Higgs are displayed in~Figure~\ref{fig:1}. We emphasise that while the first two aforementioned production mechanism have been discussed  in~\cite{Fox:2017uwr}, the third and fourth channel has not been considered in the latter paper --- the possible importance of cascades and associated $H^\pm H$ production in 2HDMs has however been stressed before in the literature~\cite{Alves:2017snd,Akeroyd:2003bt,Akeroyd:2003xi,Akeroyd:1998dt,Coleppa:2014hxa,Coleppa:2014cca,Kling:2015uba,Arhrib:2016wpw}. In~order to illustrate the four production mechanism and the resulting phenomenology, we discuss a specific benchmark scenario in each case. The discussed type-I~2HDM  benchmarks are tailored to provide explanation of other small anomalies as seen at~LEP, the Tevatron and the~LHC, while being consistent with a plethora of null~results.

\subsection{Diamond benchmark scenario}
\label{sec:bench1typeI}

\begin{figure}[!t]
\begin{center}
\includegraphics[width=0.45\textwidth]{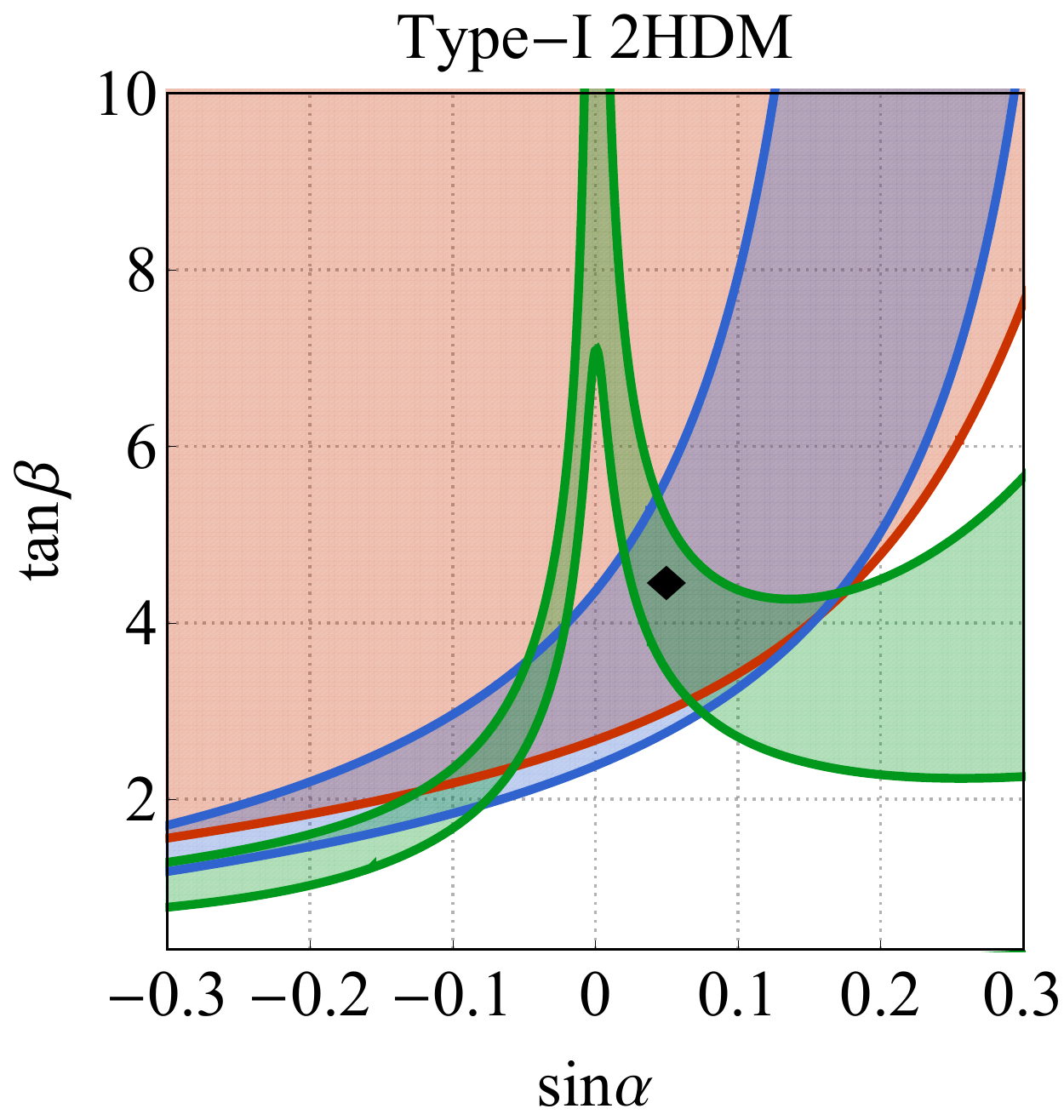}  
\vspace{2mm}
\caption{\label{fig:2} Allowed/favoured regions in the type-I~2HDM  parameter space. The  constraint shown in red is obtained from the compatibility with the LHC Run-I Higgs signal strengths~\cite{Khachatryan:2016vau}, the blue region indicates the area consistent with the LEP excess~\cite{Barate:2003sz}, while the green contour represents the parameter regions in which the di-photon $H$ signal strength  at $\sqrt{s} = 13 \, {\rm TeV}$ falls into the range of $[0.04, 0.10] \, {\rm pb}$.   The diamond ($\Diamondblack$) corresponds to the benchmark scenario~(\ref{eq:bench1typeI}). See text for further explanations.}
\end{center}
\end{figure}

The  choice of parameters  in the first type-I~2HDM benchmark scenario  is 
\begin{equation} \label{eq:bench1typeI} 
\sin \alpha  = 0.05  \,, \hspace{1mm}
\tan \beta = 4.5 \,,  \hspace{1mm}
M_H = 95 \, {\rm GeV} \,, \hspace{1mm}
M_A = 200 \, {\rm GeV} \,, \hspace{1mm}
M_{H^+} = 250 \, {\rm GeV}   \,, \hspace{1mm}
\lambda_3 = 2.3 \,.
\end{equation}

In Figure~\ref{fig:2} we show in colour the regions in the $\sin \alpha\hspace{0.5mm}$-$\hspace{0.5mm} \tan \beta$ plane that are allowed/favoured if the masses $M_H$, $M_A$, $M_{H^+}$ and the quartic coupling $\lambda_3$ are fixed to the values given in~(\ref{eq:bench1typeI}) and the mixing angles $\alpha$ and $\beta$ are varied. The red exclusion represents the $\Delta \chi^2 = 5.99$ contour (corresponding to a 95\% confidence level~(CL) for a Gaussian distribution) that follows from a $\chi^2$ analysis of the combined LHC Run-I data on Higgs production and decay rates~\cite{Khachatryan:2016vau}. Overlaid in blue is the region of parameter space corresponding to $\kappa_V^H \in [0.22, 0.38]$ that is consistent with the combined LEP data~\cite{Barate:2003sz} which shows a broad excess between $95 \, {\rm GeV}$ and $100 \, {\rm GeV}$. The green contour furthermore indicates the parameter region in which the di-photon $H$ signal strength  at $\sqrt{s} = 13 \, {\rm TeV}$ amounts to $s_H^{\gamma \gamma} = \sigma_H \, {\rm BR}_H^{\gamma \gamma} \in  [0.04, 0.10] \, {\rm pb}$. Here the shorthand  $\sigma_H = \sigma \left (pp \to H \right )$ and $ {\rm BR}_H^X = {\rm BR} \left ( H \to X \right )$ denotes the total $H$ production cross section and the  $H$ branching ratios, respectively. Values of~$s_H^{\gamma \gamma}$ in the quoted range furnish an explanation of the CMS di-photon excess~\cite{CMS-PAS-HIG-17-013}. From the location of the diamond it is evident that the benchmark scenario~(\ref{eq:bench1typeI}) accommodates the anomalies in both $e^+ e^- \to ZH$ and $pp \to H \to \gamma \gamma$, while simultaneously leading to an acceptable global Higgs fit. 

\begin{figure}[!t]
\begin{center}
\includegraphics[width=0.45\textwidth]{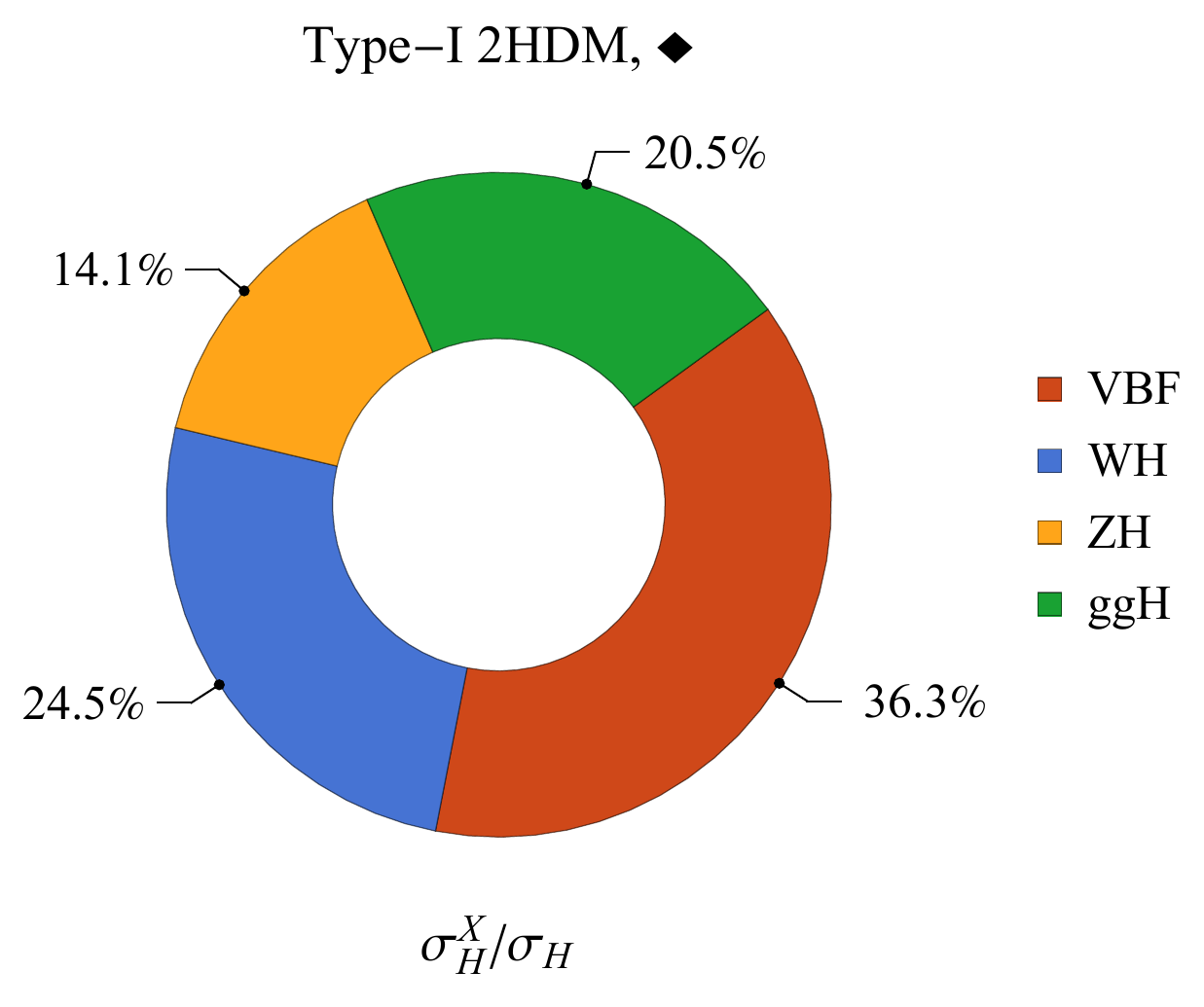}  \qquad  
\includegraphics[width=0.45\textwidth]{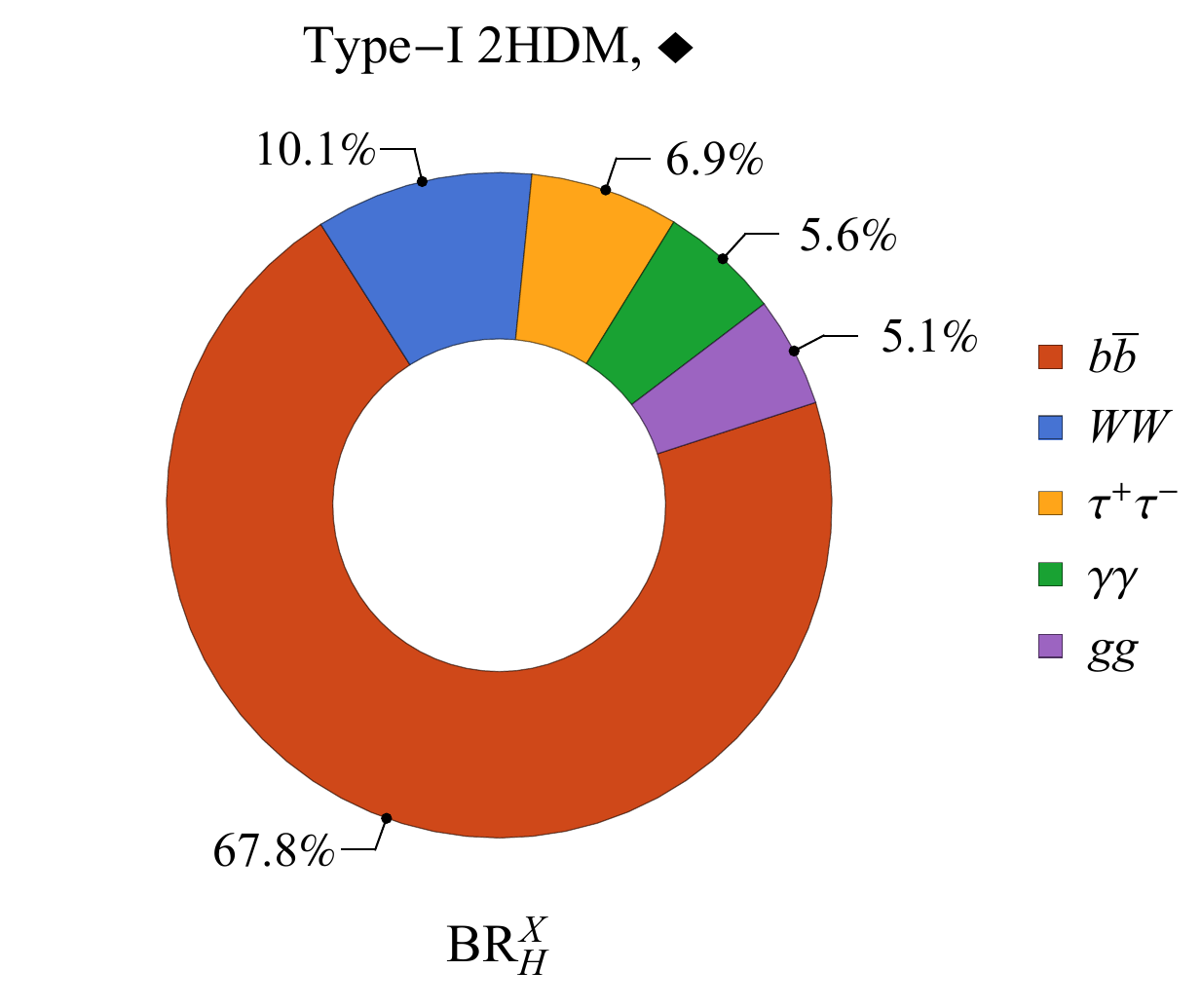}  
\vspace{2mm}
\caption{\label{fig:3}  Left: Percentage breakdown per production process of $H$ at $\sqrt{s} = 13 \, {\rm TeV}$.  Right: Branching ratios of $H$. Only $\sigma_H^X/\sigma_H$ and ${\rm BR}_H^X$ values  larger than 4\% are depicted. The shown pie charts correspond to the type-I~2HDM benchmark scenario~(\ref{eq:bench1typeI}). }
\end{center}
\end{figure}

The two panels in Figure~\ref{fig:3} show the fractional contributions $\sigma_H^X/\sigma_H$ of each channel to $H$ production~(left panel) and the branching ratios of $H$~(right panel) for the  parameters specified in~(\ref{eq:bench1typeI}). Our calculation of $\sigma_H^X$ and ${\rm BR}_H^X$ relies on the results presented in~\cite{deFlorian:2016spz} and~\cite{Djouadi:2005gi,Djouadi:2005gj}, respectively. From the left pie chart one infers that for the first benchmark scenario 74.9\% of the total cross section $\sigma_H = 1.0 \, {\rm pb}$ is due to the VBF, WH and ZH channels, while only $20.5\%$ arise from ggH production. The pie chart on the right-hand side depicts the corresponding $H$ branching ratios. We see that the five largest branching ratios are the ones to bottom quarks~($67.8\%$), $W$~bosons~($10.1\%$),  taus~($6.9\%$), photons~($5.6\%$) and gluons~($5.1\%$). The resulting signal strengths are $s_H^{b \bar b} = 0.64 \, {\rm pb}$,  $s_H^{WW} = 0.10 \, {\rm pb}$,  $s_H^{\tau^+ \tau^-} = 0.07 \, {\rm pb}$, $s_H^{\gamma \gamma} = 0.05 \, {\rm pb}$ and $s_H^{gg} = 0.05 \, {\rm pb}$.

It turns out that apart from~\cite{CMS-PAS-HIG-17-013} other existing LHC searches for neutral spin-0 resonances that probe the mass range to  $100 \, {\rm GeV}$ and below~(see~\cite{Aad:2014ioa,CMS:2016rjp,Chatrchyan:2013qga,Khachatryan:2015tra}) are not sensitive to a $H$ with such properties. To be more specific the ATLAS di-photon search~\cite{Aad:2014ioa} sets  an upper  95\%~CL limit on~$s_H^{\gamma \gamma}$ of $0.05 \, {\rm pb}$  for $M_H = 95 \, {\rm GeV}$, a factor of about 2.0 above the di-photon signal strength expected in the benchmark scenario~(\ref{eq:bench1typeI}) at $\sqrt{s} =8 \, {\rm TeV}$. The CMS search for $gg \to H \to \tau^+ \tau^-$~\cite{CMS:2016rjp} excludes values of  $s_H^{\tau^+ \tau^-}$ in excess of $34.3 \, {\rm pb}$. Compared to the di-tau signal strength given in the last paragraph  this bound is weaker by a  factor of more than  500. Searches for light Higgses in $pp \to b \bar b H X \, (H \to b \bar b)$~\cite{Chatrchyan:2013qga,Khachatryan:2015tra} are even less sensitive than the considered di-photon and di-tau analyses. We add that the decay products in $H \to b \bar b$ could in principle be reconstructed as a single, large radius high-$p_T$ jet and identified using jet substructure and dedicated $b$-tagging techniques. In fact, such a study  has been recently performed by CMS~\cite{Sirunyan:2017dgc}, observing (bounding) $Z \to b \bar b$ ($h \to b \bar b$) decays  in the single-jet topology for the first time.  However, the large $Z \to b \bar b$ background and the poor mass resolution of the reconstructed jet mass suggest that detecting the small $pp \to H \to b \bar b$ signal expected in~(\ref{eq:bench1typeI}) is impossible at the LHC even at high luminosity.

\begin{figure}[!t]
\begin{center}
\includegraphics[width=0.45\textwidth]{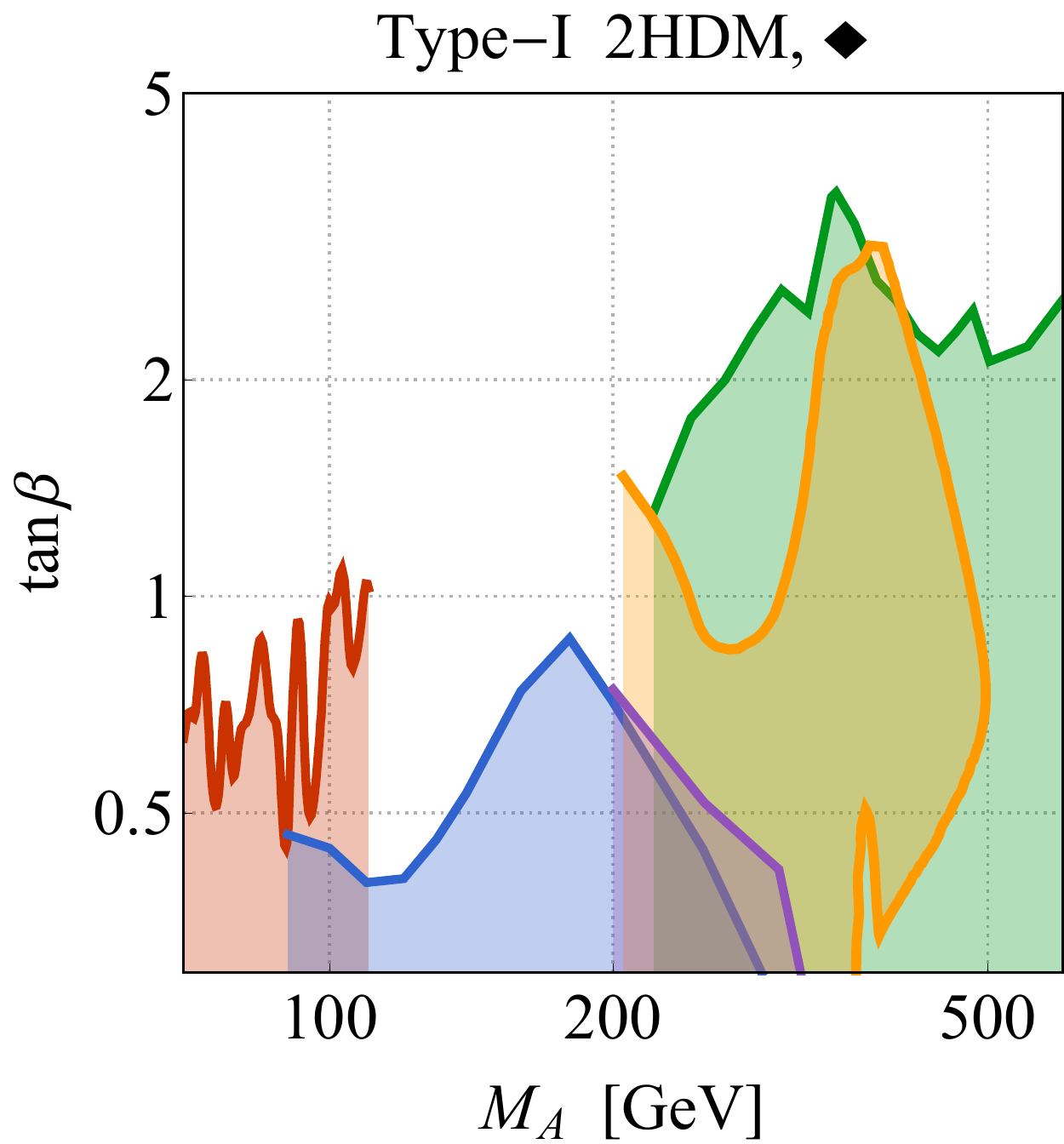} \qquad
\includegraphics[width=0.45\textwidth]{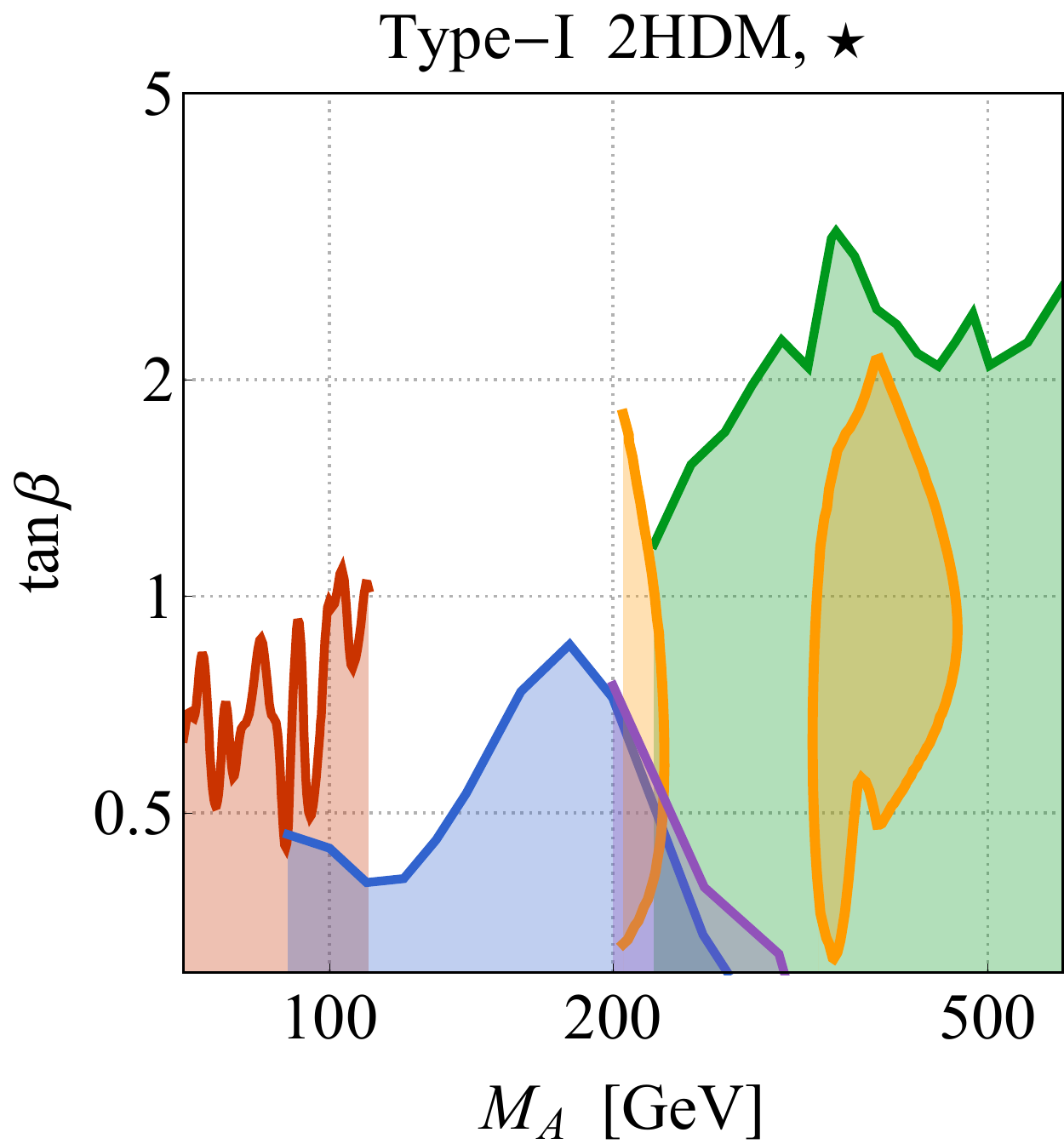}   

\vspace{6mm}

\includegraphics[width=0.45\textwidth]{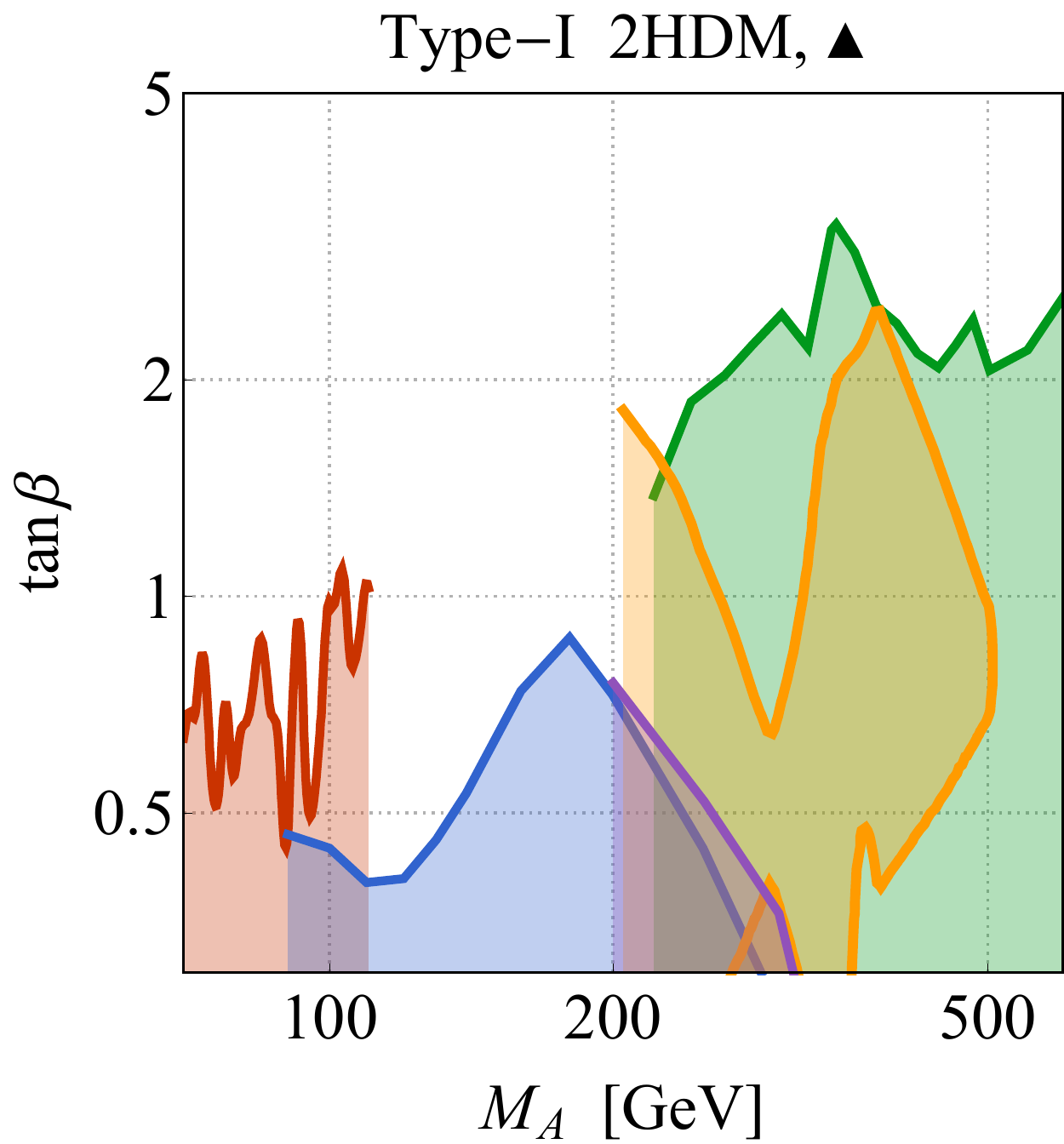} \qquad
\includegraphics[width=0.45\textwidth]{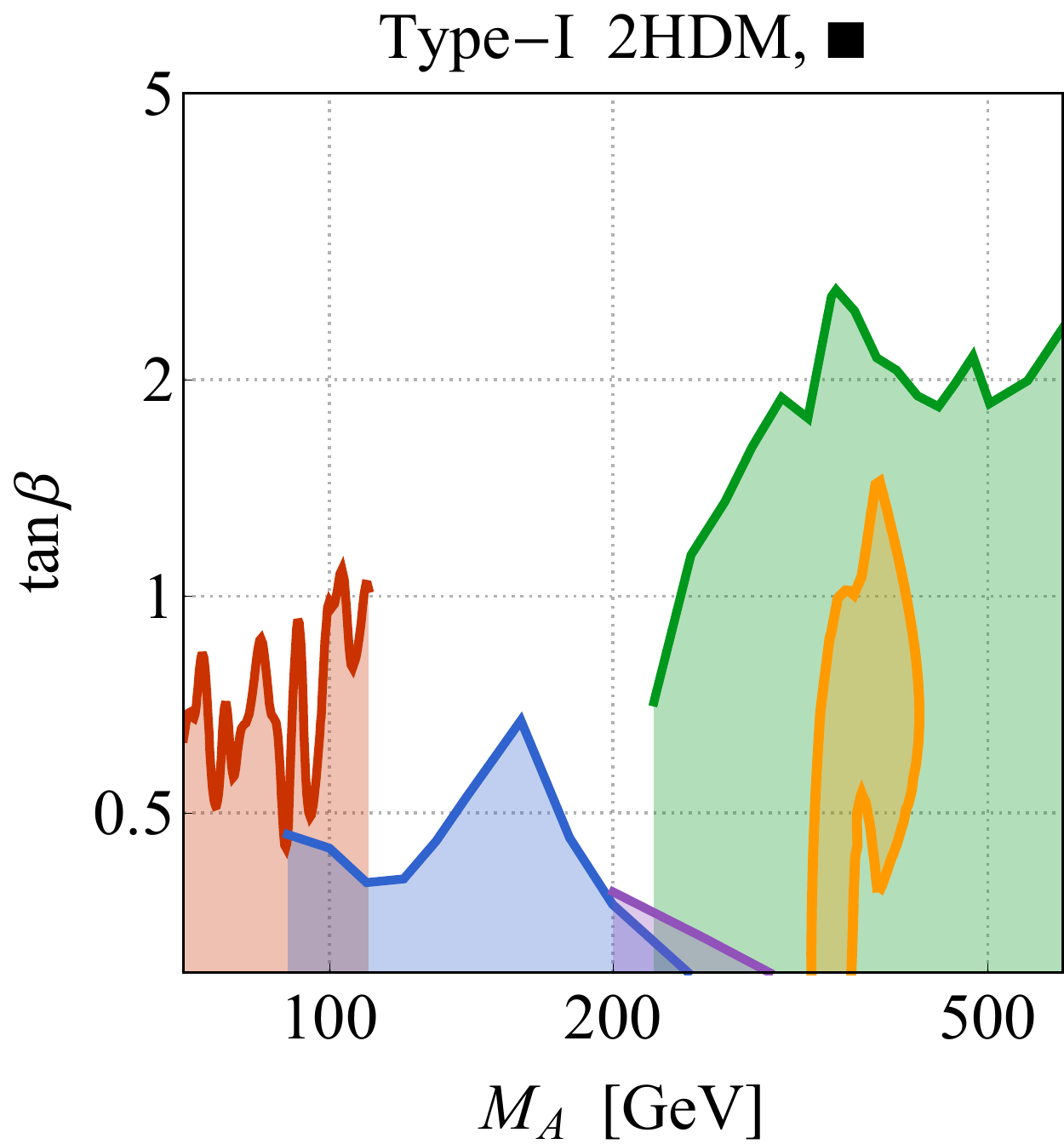}   
\vspace{2mm}
\caption{\label{fig:4a}  Constraints on the $M_A\hspace{0.25mm}$--$\hspace{0.5mm}\tan \beta$ plane in the  type-I~2HDM.  The red contours indicate the bound imposed by the $A \to \gamma \gamma$ search of CMS~\cite{CMS-PAS-HIG-17-013}, while the blue and purple parameter regions correspond to the exclusions set by the $A \to \tau^+ \tau^-$ search by CMS~\cite{CMS:2016rjp} and ATLAS~\cite{Aaboud:2017sjh}, respectively.  The green and yellow contours show furthermore the parameter sets  that are disfavoured by the ATLAS analysis of $A \to Zh$~\cite{Aaboud:2017cxo} and the CMS search for $A \to Z H$~\cite{Khachatryan:2016are}. The panels from upper left to lower right correspond to the benchmark scenarios~(\ref{eq:bench1typeI}),~(\ref{eq:bench2typeI}),~(\ref{eq:bench3typeI}) and ~(\ref{eq:bench4typeI}), respectively. All shaded regions are excluded at 95\%~CL.}
\end{center}
\end{figure}

Beside a $H$ with a mass of $95 \, {\rm GeV}$ our first type-I~2HDM benchmark scenario~(\ref{eq:bench1typeI}) also contains a relatively  light $A$ and $H^+$. The only existing LHC analyses that allow to constrain an~$A$ with a mass of $200 \, {\rm GeV}$ are the $A \to \tau^+ \tau^-$ searches~\cite{CMS:2016rjp,Aaboud:2017sjh}. The corresponding constraints are indicated in  the upper left panel of Figure~\ref{fig:4a} by the blue and purple curve, respectively.The predictions for $A$ production in $gg \to A$ have been obtained at next-to-next-to-leading order in~QCD with {\tt HIGLU}~\cite{Spira:1995mt}. One sees that for $M_A = 200 \, {\rm GeV}$  only values of $\tan \beta < 0.7$ are excluded at~95\%~CL. The benchmark scenario~(\ref{eq:bench1typeI}) however employs $\tan \beta = 4.5$ and is thus clearly allowed.

\begin{figure}[!t]
\begin{center}
\includegraphics[width=0.45\textwidth]{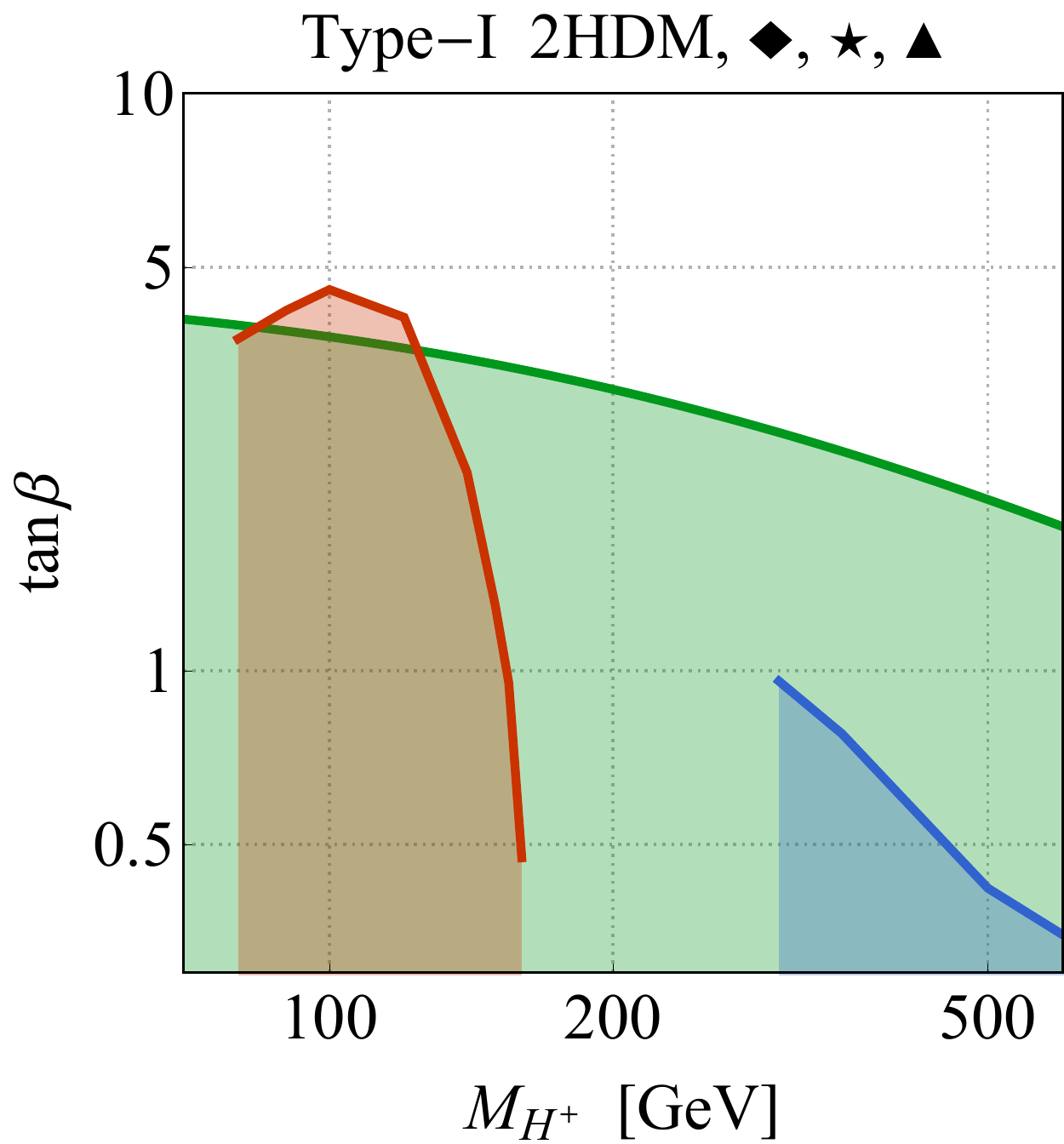} \qquad
\includegraphics[width=0.45\textwidth]{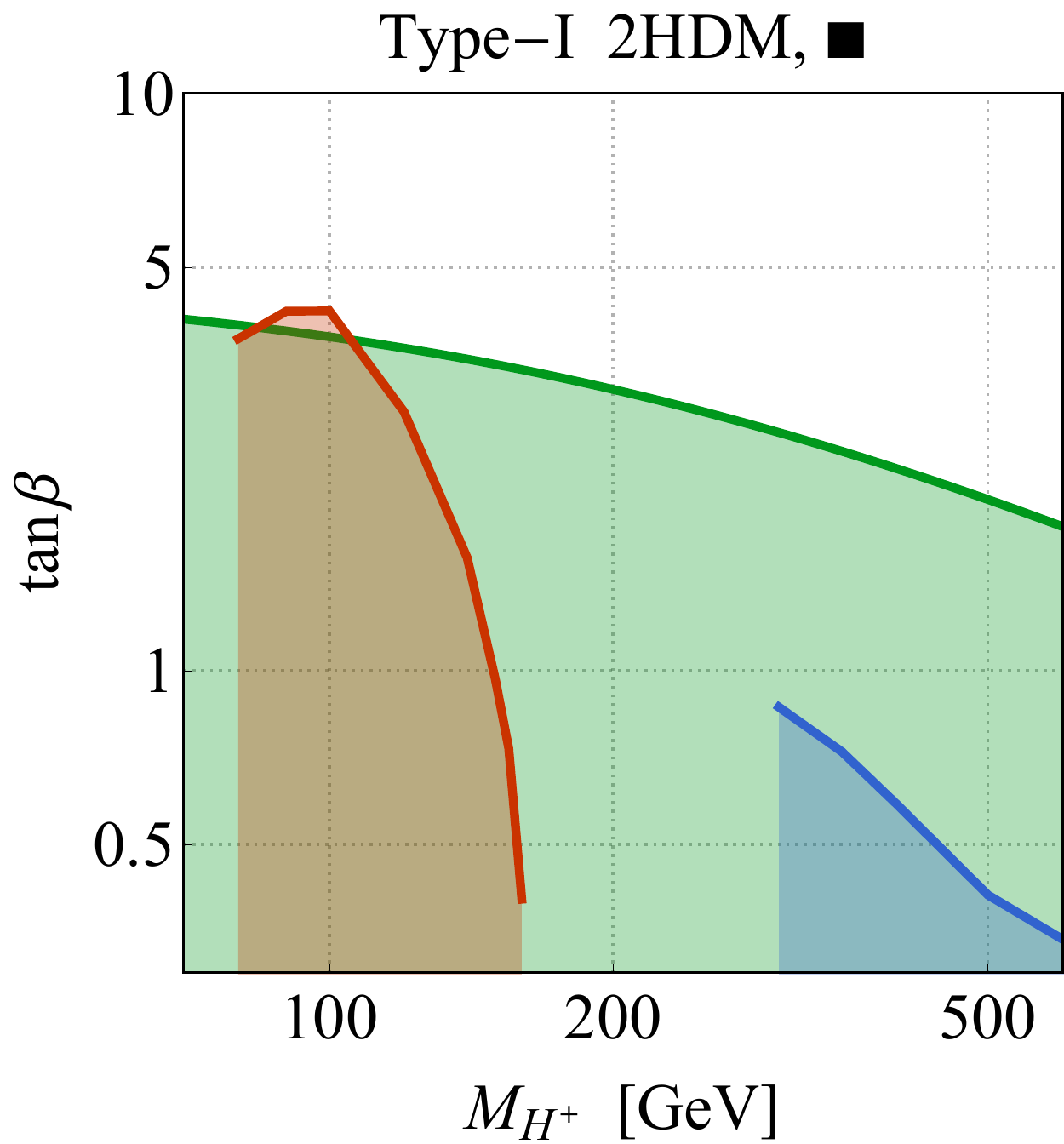}   
\vspace{2mm}
\caption{\label{fig:4b}  Constraints on the $M_{H^+}\hspace{0.25mm}$--$\hspace{0.5mm}\tan \beta$ plane in the  type-I~2HDM. The red exclusions are obtained from the CMS~$H^+ \to \tau^+  \nu_\tau$ search~\cite{CMS-PAS-HIG-16-031}, while the blue contours indicate the limits set by the ATLAS search for $H^+ \to t \bar b$~\cite{ATLAS-CONF-2016-089}.  Overlaid in green is the parameter space disfavoured by the $B_s \to \mu^+ \mu^-$ measurements of CMS and LHCb~\cite{CMS:2014xfa,Aaij:2017vad}. The limits in the benchmark scenarios~(\ref{eq:bench1typeI}),~(\ref{eq:bench2typeI}) and ~(\ref{eq:bench3typeI}) are almost identical and shown on the left, whereas the bounds that are relevant in the case of~(\ref{eq:bench4typeI}) are displayed on the right. In both panels the shaded regions are excluded at 95\%~CL.}
\end{center}
\end{figure}

Direct limits on charged Higgs masses  above the top threshold are due to the LHC searches for  $H^+ \to t \bar b$ (for the latest $\sqrt{s}= 13 \, {\rm TeV}$ analysis see~\cite{ATLAS-CONF-2016-089}) while indirect constraints on $M_{H^+}$ are provided by $B \to X_s \gamma$~\cite{Hermann:2012fc,Misiak:2015xwa,Misiak:2017bgg}, $B$-meson mixing~\cite{Abbott:1979dt,Geng:1988bq,Buras:1989ui,Kirk:1634844} as well as  $B_s \to \mu^+ \mu^-$~\cite{Chankowski:2000ng,Bobeth:2013uxa,CMS:2014xfa,Aaij:2017vad}, but also follow from $Z \to b \bar b$~\cite{Denner:1991ie,Haisch:2007ia,Freitas:2012sy} and the $\rho$ parameter (the relevant formulas can be found in~\cite{Djouadi:2005gj}  for instance).  The most stringent constraints on the $M_{H^+}\hspace{0.25mm}$--$\hspace{0.5mm}\tan \beta$ plane for the case of the type-I~2HDM are summarised in Figure~\ref{fig:4b}. The results for the $H^+$ production cross sections are taken from~\cite{deFlorian:2016spz}. The constraints that apply in the case of~(\ref{eq:bench1typeI}) are shown in the left panel of the figure. One observes that  the measurements of $B_s \to \mu^+ \mu^-$~\cite{CMS:2014xfa,Aaij:2017vad} provide at present the strongest  limits on $\tan \beta$ for most  charged Higgs masses. Numerically, we find for $M_{H^+} = 250 \, {\rm GeV}$ the bound $\tan \beta > 2.8$, which does not rule out the choice of $\tan \beta$ made in~(\ref{eq:bench1typeI}).  Improved LHC searches for $B_s \to \mu^+ \mu^-$ should however be able to exclude or find evidence for scenarios with $\tan \beta \lesssim 4$ and a charged Higgs with a mass not too far above the top threshold.

Since the charged Higgses couple to the CP-even spin-0 states, a lightish $H^+$ in general also modifies $\Gamma \left (h \to \gamma \gamma \right )$ and $\Gamma \left (H \to \gamma \gamma \right )$. The  size of the modifications is however  model dependent, because the form of the trilinear couplings $\lambda_{hH^+ H^-}$ and $\lambda_{HH^+ H^-}$  depends  sensitively on the choice of the scalar potential. For our potential~(\ref{eq:VH}) it is always possible to arrange for  the parameter $\kappa_\gamma^h = \sqrt{\Gamma \left ( h \to \gamma \gamma \right )/\Gamma \left ( h \to \gamma \gamma  \right )_{\rm SM}}$ to agree with the LHC Run-I result of $\kappa_\gamma^h = 0.87^{+0.14}_{-0.09}$~\cite{Khachatryan:2016vau} by tuning $\lambda_3$ for any given set of $\alpha$, $\beta$ and $M_{H^+}$. The parameters~(\ref{eq:bench1typeI})  in fact lead to~$\kappa^h_{\gamma} = 0.78$, and we find that charged Higgs loops suppress  $\Gamma \left (h \to \gamma \gamma \right )$ by around~15\% with respect to the case when only top and $W$-boson loops are considered. Since the effects of charged Higgs loops are non-negligible for the parameter choices~(\ref{eq:bench1typeI}), we have, unlike~\cite{Fox:2017uwr}, included them in~Figure~\ref{fig:2} and in the right pie chart of Figure~\ref{fig:3}. We furthermore note that  that  for the adopted values of $M_H$, $M_A$, $M_{H^+}$ and $\lambda_3$, one can show~(cf.~\cite{Eberhardt:2013uba}) that the resulting Higgs potential~(\ref{eq:VH}) is bounded from below and that the constraints arising from the $\rho$ parameter are satisfied within~$2\sigma$,~i.e.~the value of $\Delta \rho = \rho - 1$ falls into the range $[-1.2, 2.4 ] \cdot 10^{-3}$~\cite{Patrignani:2016xqp}.

\subsection{Star benchmark scenario}
\label{sec:bench2typeI}

\begin{figure}[!t]
\begin{center}
\includegraphics[width=0.45\textwidth]{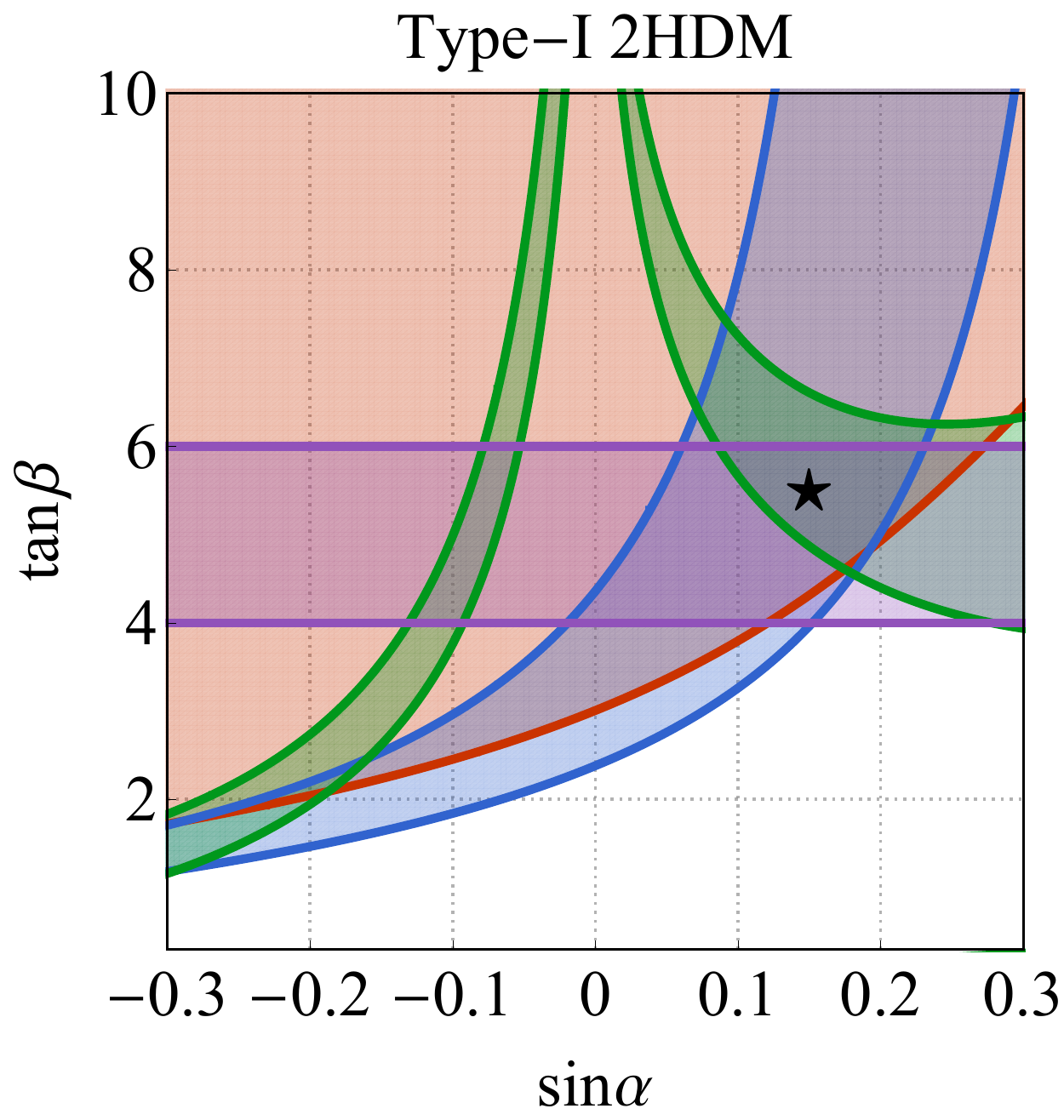}  
\vspace{2mm}
\caption{\label{fig:5}  As Figure~\ref{fig:2} but for the star ($\star$) benchmark scenario~(\ref{eq:bench2typeI}). The range of $\tan \beta$ favoured by  the leptonic excess in $t \bar t h$ production~\cite{ATLAS-CONF-2016-058} is indicated in purple. See text for further details.}
\end{center}
\end{figure}

The second type-I~2HDM benchmark scenario that we study in detail is defined by
\begin{eqnarray}  \label{eq:bench2typeI} 
\sin \alpha  = 0.15 \,, \hspace{1mm}
\tan \beta = 5.5 \,,  \hspace{1mm}
M_H = 95 \, {\rm GeV} \,, \hspace{1mm}
M_A = 205\, {\rm GeV} \,, \hspace{1mm}
M_{H^+} = 125 \, {\rm GeV}  \,, \hspace{1mm}
\lambda_3 = 0.55 \,. \hspace{4mm}
\end{eqnarray}

The constraints on this benchmark scenario following from a global fit to the LHC Run-I Higgs data~(red), the region favoured by the LEP anomaly in $e^+ e^- \to ZH$~(blue) as well as the di-photon excess observed at CMS~(green) are shown in Figure~\ref{fig:5}.  The horizontal band~(purple) corresponds to  $\tan \beta$ values in the range of $[4,6]$, which have been shown in~\cite{Alves:2017snd} to be favoured by the leptonic excess in $t \bar t h$  production as seen by ATLAS~in the $\sqrt{s} = 13 \, {\rm TeV}$ data~\cite{ATLAS-CONF-2016-058}. The depicted constraints are obtained by fixing $M_H$, $M_A$, $M_{H^+}$ and $\lambda_3$ to the values quoted above and varying $\alpha$ and $\beta$. The star indicates the choice of $\sin \alpha$ and $\tan \beta$ made in~(\ref{eq:bench2typeI}). Since it is located in the overlap of all four shaded regions, it is not only consistent with the combined LHC Run-I Higgs data, but at the same time also fits  the deviations seen in $e^+ e^- \to ZH$, $pp \to H \to \gamma \gamma$ and $pp \to t \bar t h$.  

\begin{figure}[!t]
\begin{center}
\includegraphics[width=0.45\textwidth]{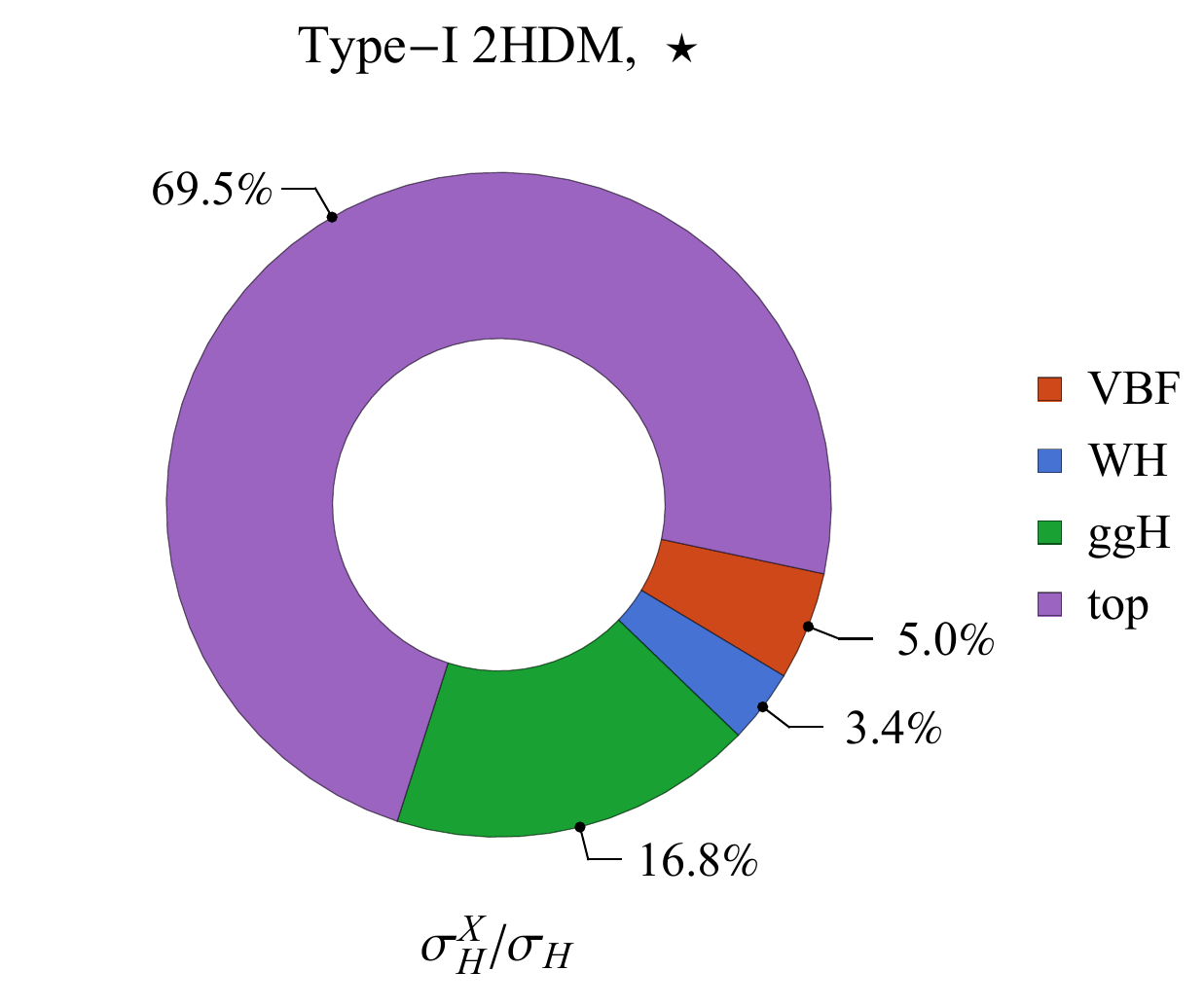}  \qquad  
\includegraphics[width=0.45\textwidth]{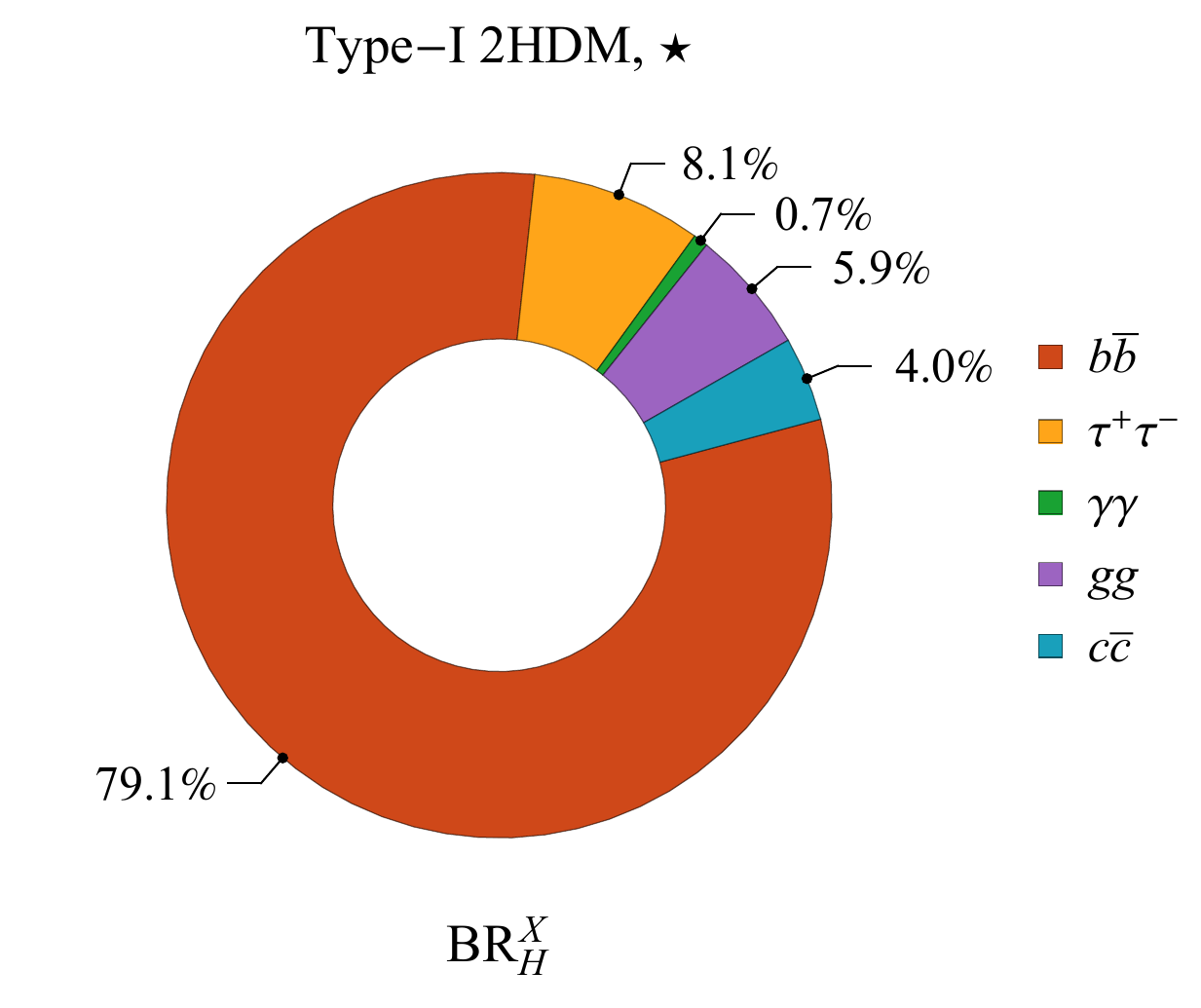}  
\vspace{2mm}
\caption{\label{fig:6} As Figure~\ref{fig:3} but for the second type-I~2HDM   benchmark scenario~(\ref{eq:bench2typeI}).  The label ``top'' in the left panel refers to $H$ production via the cascade $\overset{\text{\relsize{-2}(}-\text{\relsize{-2})}}{t} \to H^{\pm} \overset{\text{\relsize{-2}(}-\text{\relsize{-2})}}{b} \to W^{\pm \, \ast} H \overset{\text{\relsize{-2}(}-\text{\relsize{-2})}}{b}$ starting from top-quark pair or single-top events. With the exception of ${\rm BR}_H^{\gamma \gamma}$ only  values of  $\sigma_H^X/\sigma_H$ and ${\rm BR}_H^X$ larger than 3\% are shown. }
\end{center}
\end{figure}

Figure~\ref{fig:6} illustrates the importance of the different $H$ production channels~(left panel) and the values of the branching ratios of $H$~(right panel) for the type-I~2HDM parameter scenario~(\ref{eq:bench2typeI}).  From the left pie chart one observes that only $8.4\%$ ($16.8\%$) of $\sigma_H = 10.5 \, {\rm pb}$ is due to the combination of the VBF and WH  modes (the ggH channel), while the bulk of the total cross section of $69.5\%$ arises from top-pair and single-top production with the top or anti-top cascading to a $H$ through $\overset{\text{\relsize{-2}(}-\text{\relsize{-2})}}{t} \to H^{\pm} \overset{\text{\relsize{-2}(}-\text{\relsize{-2})}}{b} \to W^{\pm \, \ast} H \overset{\text{\relsize{-2}(}-\text{\relsize{-2})}}{b}$ --- see left diagram in Figure~\ref{fig:1}. In our numerical analysis, we employ the values $829 \, {\rm pb}$~\cite{Czakon:2013goa}   and $288\, {\rm pb}$~\cite{Kant:2014oha}  for the top-pair and single-top production cross section, respectively. These numbers correspond to $pp$ collisions at $\sqrt{s} = 13 \, {\rm TeV}$. 

\begin{figure}[!t]
\begin{center}
\includegraphics[width=0.45\textwidth]{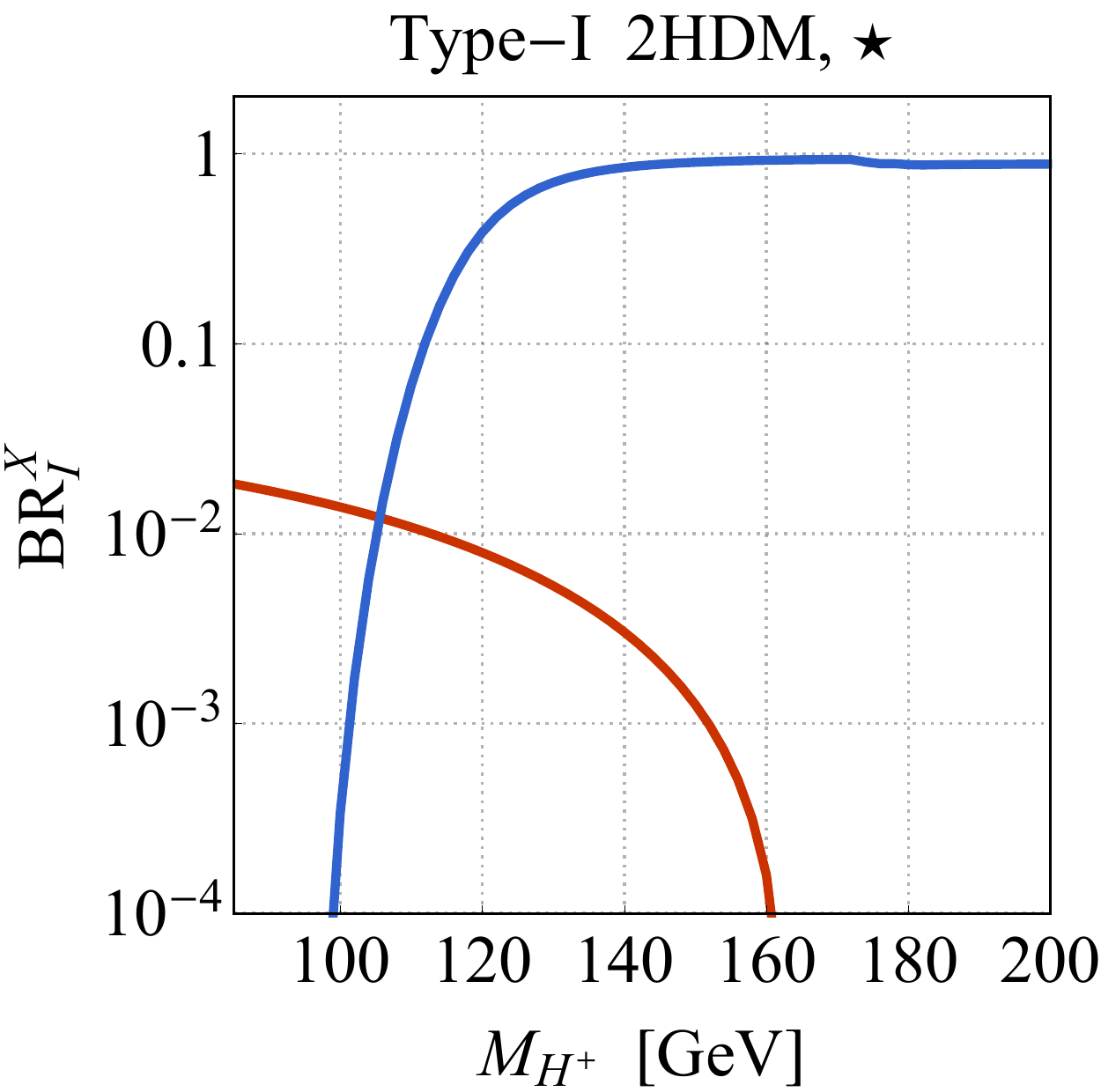}  \qquad  
\includegraphics[width=0.45\textwidth]{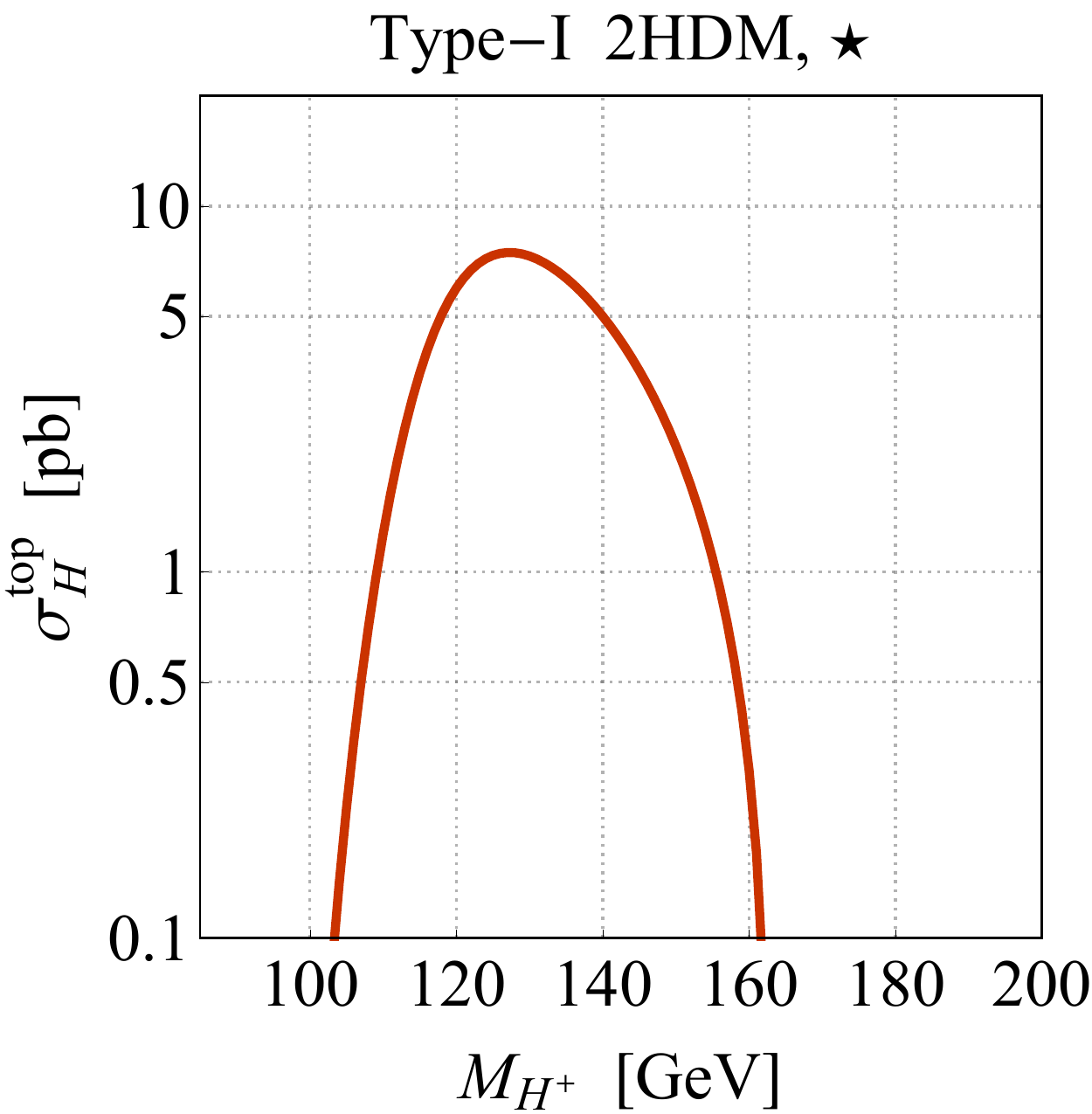}  
\vspace{2mm}
\caption{\label{fig:7} Left: The  $t \to H^+ b$~(red) and $H^+ \to W^{+} H$~(blue) branching ratio as a function of $M_{H^+}$. Right:~The production cross section of $H$ via top-pair and single-top production at $\sqrt{s} = 13 \, {\rm TeV}$ multiplied by the branching ratios for $\overset{\text{\relsize{-2}(}-\text{\relsize{-2})}}{t} \to H^{\pm} \overset{\text{\relsize{-2}(}-\text{\relsize{-2})}}{b} \to W^{\pm \, \ast} H \overset{\text{\relsize{-2}(}-\text{\relsize{-2})}}{b}$. Both panels show results  in the benchmark scenario~(\ref{eq:bench2typeI}).}
\end{center}
\end{figure}

The dominance of cascade $H$ production for the parameter choices~(\ref{eq:bench2typeI}) is easy to understand by analysing the $M_{H^+}$-dependence of ${\rm BR}_t^X$ and ${\rm BR}_{H^+}^X$. We show the relevant branching ratios in the left panel of Figure~\ref{fig:7}. Our calculation of the branching ratios is based on the formulas given in~\cite{Djouadi:2005gj,Alves:2017snd,Djouadi:1995gv}. One observes that while ${\rm BR}_t^{H^+ b}$ decreases from around $1\%$ to $0.1\%$ between  $M_{H^+} = 110 \, {\rm GeV}$ and $M_{H^+} = 155 \, {\rm GeV}$, the branching ratio ${\rm BR}_{H^+}^{W^{+ \, \ast} H}$ simultaneously increases  from roughly $5\%$ to $90\%$. As a result one obtains $H$ production cross sections of $\sigma_H^{\rm top} \gtrsim 1 \, {\rm pb}$  for  $M_{H^+} \underset{\text{\relsize{+2}$\tilde{}$}}{\in} [110, 155] \, {\rm GeV}$.  This feature is illustrated on the right in Figure~\ref{fig:7}. From the right panel in Figure~\ref{fig:6} one furthermore observes that the three largest branching ratios in our second benchmark scenario~(\ref{eq:bench2typeI}) are the ones to bottom pairs, taus and gluons. These channels amount to $79.1\%$, $8.1\%$ and $5.9\%$ of the total decay width of $H$, while the di-photon branching ratio constitutes just a mere fraction of~$0.7\%$. The corresponding signal strengths amount to  $s_H^{b \bar b} = 8.3 \, {\rm pb}$,  $s_H^{\tau^+ \tau^-} = 0.85 \, {\rm pb}$,  $s_H^{gg} = 0.62 \, {\rm pb}$ and  $s_H^{\gamma \gamma} = 0.07 \, {\rm pb}$ at $\sqrt{s} = 13 \, {\rm TeV}$. At $\sqrt{s} = 8 \, {\rm TeV}$ we find that the di-photon signal strength of $H$ is a factor of around two below the sensitivity of the ATLAS search~\cite{Aad:2014ioa}.

Since  in the benchmark scenario~(\ref{eq:bench2typeI}) the $95 \, {\rm GeV}$ Higgs $H$ is produced dominantly in association with top quarks, a couple of comments concerning the  Tevatron and LHC searches that target final states of this type seem to be in order. The existing $t \bar t h$ searches fall broadly speaking into two classes. Firstly, more exclusive analyses~(see~\cite{CMS-PAS-HIG-16-038,ATLAS-CONF-2016-067,CMS-PAS-HIG-16-020,CMS-PAS-HIG-17-004} for the latest LHC searches of this type) that employ multivariate discriminants such as boosted decision decisions trees or neural networks, and are specifically tuned to the final state kinematics of  the SM signal. Second, more inclusive searches based on cut-and-count approaches that impose only  rather loose selection requirements to suppress backgrounds. Examples of the second type are the CDF searches for $t \bar t h \, ( h \to b \bar b)$~\cite{Collaboration:2012bk,Aaltonen:2013ipa} and the ATLAS~\cite{ATLAS-CONF-2016-058} and CMS~\cite{CMS-PAS-HIG-16-022} analyses that both look for associated production of a Higgs boson and $t \bar t$ in multi-lepton final states. The interesting observation is now that while most of the exclusive $t \bar t h$  analyses show no significant deviations from the SM expectations or are inconclusive, the aforementioned  inclusive results display small excesses. In fact, it has been pointed out in~\cite{Alves:2017snd} that the existing excesses in $t \bar t h$ searches can be explained by the contamination from $\overset{\text{\relsize{-2}(}-\text{\relsize{-2})}}{t} \to H^{\pm} \overset{\text{\relsize{-2}(}-\text{\relsize{-2})}}{b} \to W^{\pm \, \ast} H \overset{\text{\relsize{-2}(}-\text{\relsize{-2})}}{b}$ followed by $H \to b \bar b, \tau^+ \tau^-$, and that a model that naturally leads to such a contamination is the type-I~2HDM with low to moderate $\tan \beta$ and a light Higgs spectrum. The~$\tan \beta$ range  of~$[4,6]$ that is favoured by the ATLAS multi-lepton excess in $t \bar t h$~\cite{Alves:2017snd} is indicated in  Figure~\ref{fig:5} by a purple stripe. Concerning the latest combined $h \to \gamma \gamma$ measurements by ATLAS~\cite{ATLAS-CONF-2016-067} and CMS~\cite{CMS-PAS-HIG-16-020} it is important to mention that these analyses include the $t \bar t h$ channel, but would have barely missed a $H$ with $95 \, {\rm GeV}$, because they only considered  di-photon invariant masses $m_{\gamma \gamma} \in [105, 160] \, {\rm GeV}$ and $m_{\gamma \gamma} \in [100, 180] \, {\rm GeV}$, respectively. Future LHC searches for $t \bar t H \, (H \to \gamma \gamma)$ with an enlarged mass window should however find clear evidence of a signal, if the $95 \, {\rm GeV}$ di-photon excess is a true sign of new physics and not just a fluke. 

\begin{figure}[!t]
\begin{center}
\includegraphics[width=0.45\textwidth]{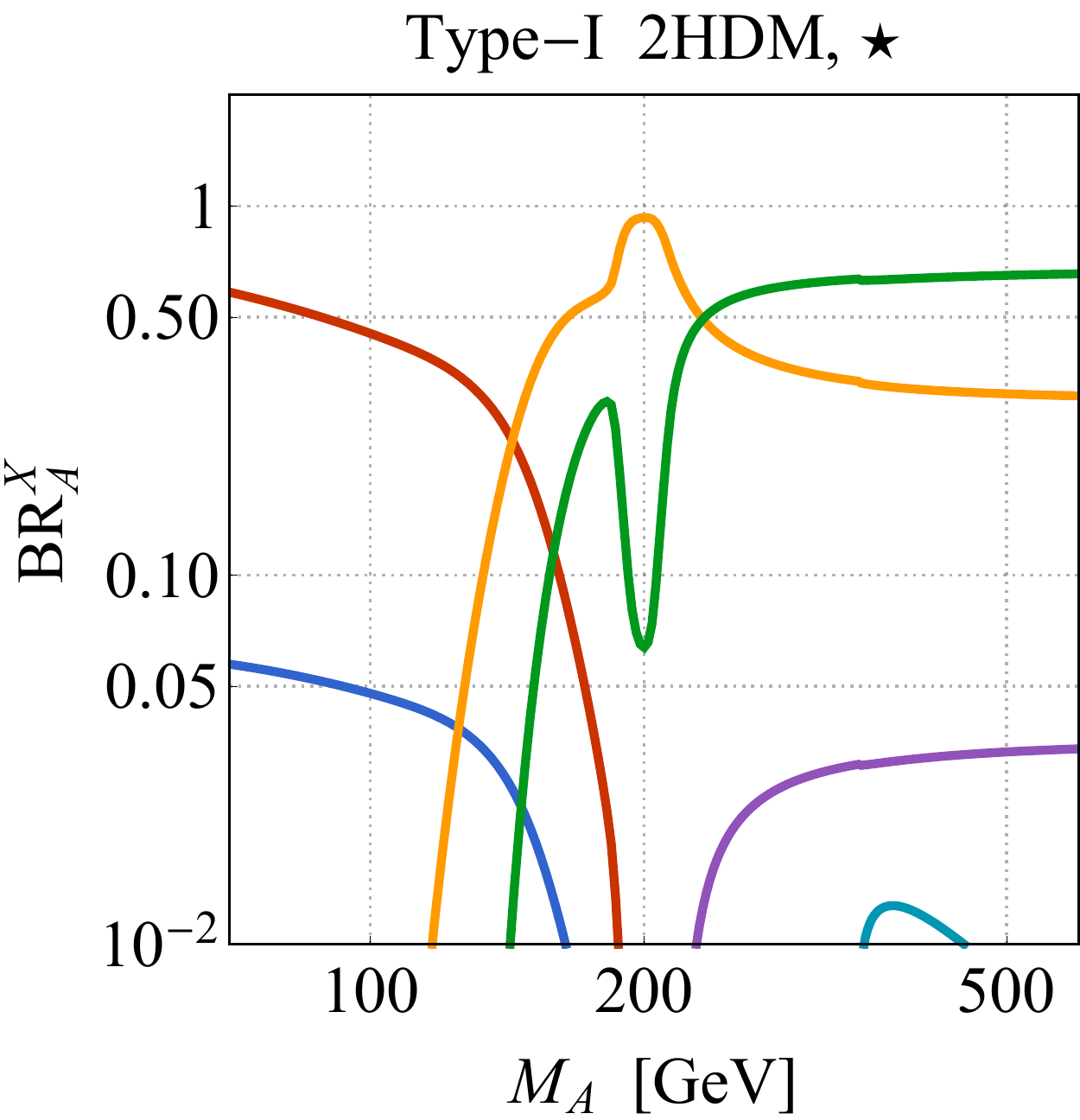}  \qquad  
\includegraphics[width=0.45\textwidth]{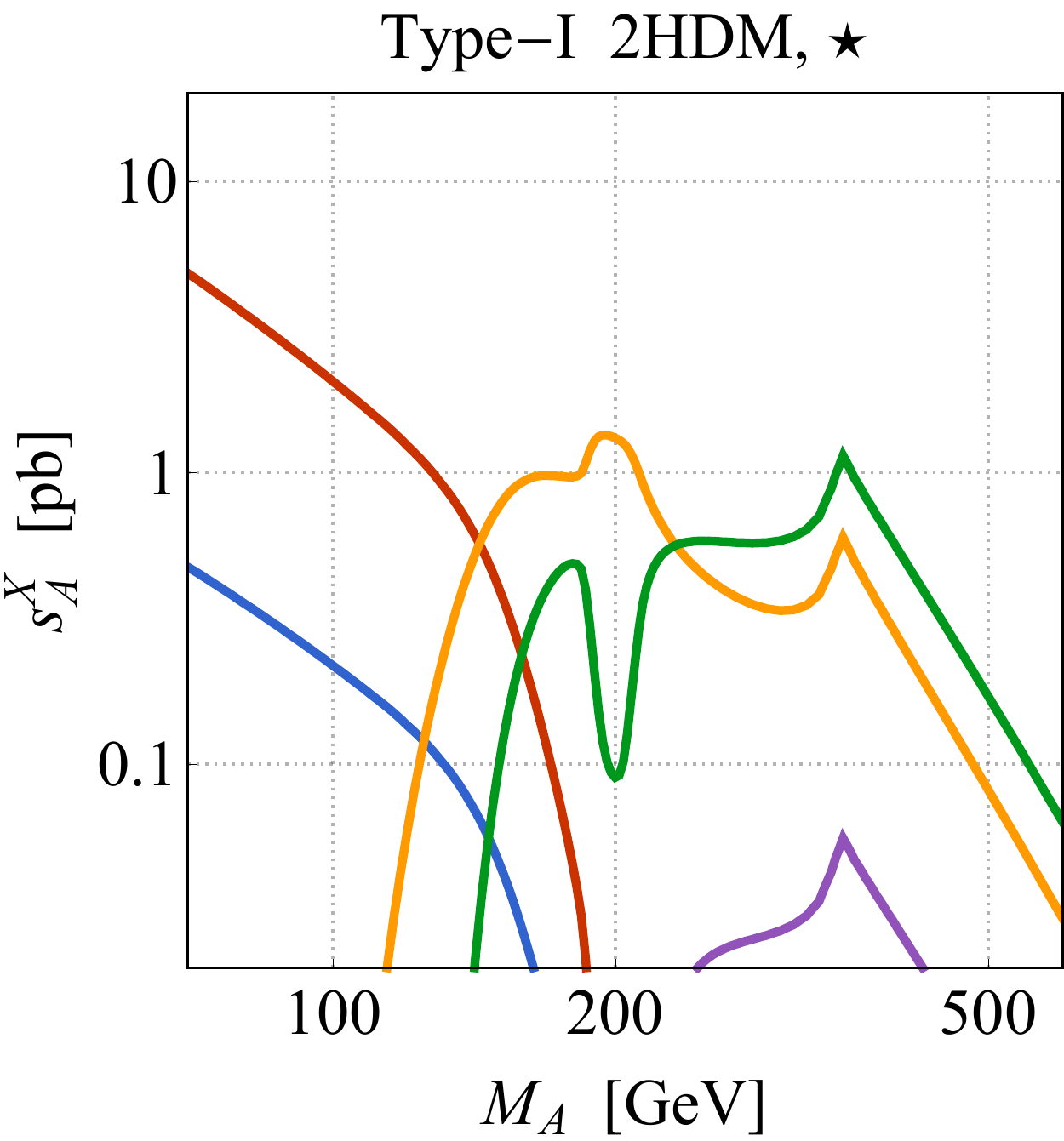}  
\vspace{2mm}
\caption{\label{fig:8} Left: Branching ratios of $A$ into $b \bar b$~(red), $\tau^+ \tau^-$~(blue), $ZH$~(yellow), $W^{\mp} H^{\pm}$~(green), $Zh$~(purple) and $t \bar t$~(cyan). Right: Signal strengths for $gg \to A \to X$ at $\sqrt{s} = 13 \, {\rm TeV}$. The final states $X$ are $b \bar b$~(red), $\tau^+ \tau^-$~(blue), $ZH$~(yellow), $W^{\mp} H^{\pm}$~(green), $Zh$~(purple) and $t \bar t$~(cyan). All results correspond to the benchmark scenario~(\ref{eq:bench2typeI}).}
\end{center}
\end{figure}

The Higgs spectrum of~(\ref{eq:bench2typeI}) also contains a lightish $A$ with a mass of $205 \, {\rm GeV}$. In order to understand how to search for such a pseudoscalar in the most efficient way, we show in the two panels of Figure~\ref{fig:8} the branching ratios of $A$ (left) and the corresponding $gg \to A \to X$ signal strengths at $\sqrt{s} = 13 \, {\rm TeV}$~(right). One observes that  for $M_A \gtrsim 160 \, {\rm GeV}$ ($M_A \gtrsim 135 \, {\rm GeV}$) the $A \to ZH$  branching ratio exceeds the one to bottom (tau) pairs. In consequence, LHC searches for  $ZH$  production with $Z \to \ell^+ \ell^-$ and $H \to b \bar b, \tau^+ \tau^-$~\cite{Khachatryan:2016are,CMS-PAS-HIG-16-010} provide good opportunities to test and to constrain type-I~2HDM realisation with a neutral Higgs spectrum {\`a} la~(\ref{eq:bench2typeI}).  Although it is parametrically suppressed by a factor of $\cot^2 \left ( \beta - \alpha \right )$ compared to $A \to ZH$, another interesting probe of such fermiophobic scenarios is the $A  \to Zh$ channel~(see~e.g.~\cite{ATLAS:2017nxi}). In fact, as can be seen from Figure~\ref{fig:4a}, for $M_A > 205 \, {\rm GeV}$ the existing searches for $A \to Zh/H$ provide the most stringent bounds on $\tan \beta$ in the case of all benchmark scenarios. For $M_A = 205 \, {\rm GeV}$, we find that the parameter space with $\tan \beta < 1.8$ is excluded at 95\%~CL by the  CMS search for $A \to ZH$~\cite{Khachatryan:2016are}. The benchmark scenario~(\ref{eq:bench2typeI}) is thus clearly viable. Notice that the limit on $\tan \beta$ that we have derived from the $A \to ZH$ search ends slightly above $200 \, {\rm GeV}$, because the CMS collaboration studies only signal benchmarks with $M_A > M_H + M_Z$. Since off-shell decays of $A$ to $ZH$ are important in our case (see left panel in Figure~\ref{fig:8}) dropping this restriction would allow to extend the shown bound down to $M_A < M_H + M_Z$. Given this limitation and the fact that~\cite{Khachatryan:2016are}  is based on only $19.8 \, {\rm fb}^{-1}$ of $\sqrt{s} = 8 \, {\rm TeV}$ data, one can expect future LHC searches for $A \to ZH$ to be able to notably improve the constraints on fermiophobic~type-I~2HDM scenarios.   We finally add  that  the parameter choices~(\ref{eq:bench2typeI}) give rise to a signal strength  of  around $66 \, {\rm fb}$~($7 \, {\rm fb}$) for $pp \to A \to ZH \, (Z \to \ell^+ \ell^-)$ in the $H \to b \bar b$ ($H \to \tau^+ \tau^-$) channel at $\sqrt{s} = 13 \, {\rm TeV}$. 

The most relevant constraints on the $M_{H^+}\hspace{0.25mm}$--$\hspace{0.5mm}\tan \beta$ plane for the case of the type-I~2HDM are shown in Figure~\ref{fig:4b}.  For $M_{H^+} = 125 \, {\rm GeV}$ one observes from the left panel that values of $\tan \beta < 3.7$ are disfavoured at 95\%~CL by the latest CMS search for $H^+ \to \tau^+ \nu_\tau$~\cite{CMS-PAS-HIG-16-031}. The choice of $\tan \beta = 5.5$ made in~(\ref{eq:bench2typeI}) represents  therefore a viable option. We also emphasise that the contributions of charged Higgs loops to $\Gamma \left (h \to \gamma \gamma \right )$ and $\Gamma \left (H \to \gamma \gamma \right )$ have been taken into account  in~Figure~\ref{fig:5} and in the pie chart shown on the right-hand side  in Figure~\ref{fig:6}.  Numerically, we find that $\kappa_\gamma^h = 0.80$ and observe that charged Higgs effects in the benchmark scenario~(\ref{eq:bench2typeI})  suppress the~$h$~($H$) di-photon decay rate by around $10\%$ ($30\%$).  The choices of $M_H$, $M_A$, $M_{H^+}$ and~$\lambda_3$ employed in~(\ref{eq:bench2typeI}) finally lead to a Higgs potential~(\ref{eq:VH}) that is bounded from below and to a $\rho$ parameter that is compatible with the existing $2\sigma$ limits. 

\subsection{Triangle benchmark scenario}
\label{sec:bench3typeI}

In our third type-I~2HDM benchmark scenario  we adopt the following choice of parameters 
\begin{equation} \label{eq:bench3typeI}
\sin \alpha  = 0.1  \,, \hspace{1mm}
\tan \beta = 4 \,,  \hspace{1mm}
M_H = 95 \, {\rm GeV} \,, \hspace{1mm}
M_A = 350 \, {\rm GeV} \,, \hspace{1mm}
M_{H^+} = 170 \, {\rm GeV}   \,, \hspace{1mm}
\lambda_3 = 0.9 \,. 
\end{equation}

\begin{figure}[!t]
\begin{center}
\includegraphics[width=0.45\textwidth]{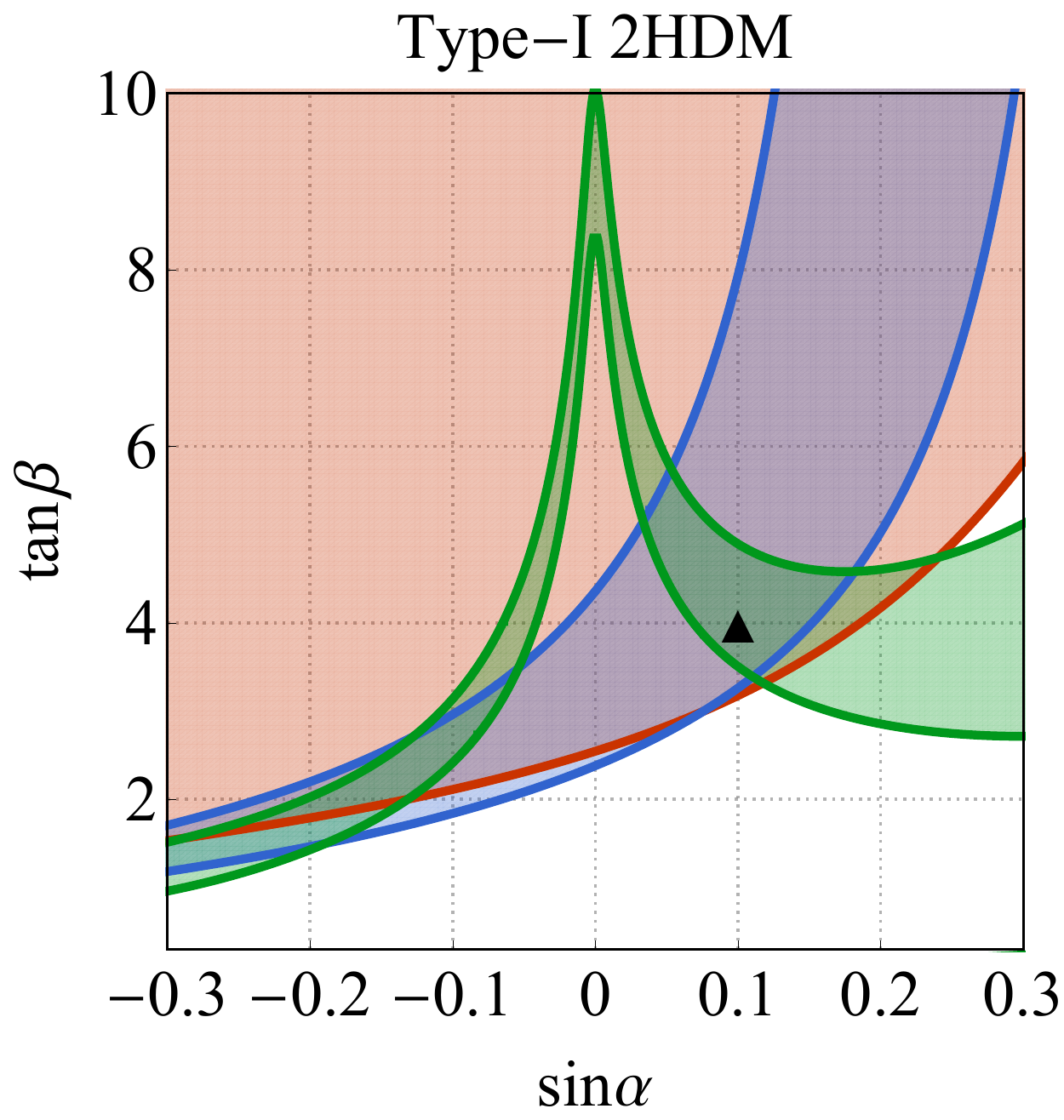}  
\vspace{2mm}
\caption{\label{fig:9}  Constraints on the type-I~2HDM in the case of the triangle ($\blacktriangle$) benchmark scenario~(\ref{eq:bench3typeI}). The colour coding resembles the one of Figure~\ref{fig:2}.}
\end{center}
\end{figure}

The constraints on this benchmark scenario are summarised in Figure~\ref{fig:9}. The region preferred by a global analysis of the LHC Run-I Higgs data is coloured red, while the regions favoured by the LEP excess in $e^+ e^- \to ZH$ and the CMS di-photon anomaly are indicated in blue and green. The displayed constraints are obtained by setting $M_H$, $M_A$, $M_{H^+}$ and $\lambda_3$ to the values reported in~(\ref{eq:bench3typeI}), while $\alpha$ and $\beta$ are left to vary. The triangle marks the values of $\sin \alpha$ and $\tan \beta$ chosen in the benchmark scenario. As is evident from the figure, the choices~(\ref{eq:bench3typeI}) also lead to an explanation of both the LEP and CMS anomalies, while not being in conflict with LHC~Run-I Higgs data. 

\begin{figure}[!t]
\begin{center}
\includegraphics[width=0.45\textwidth]{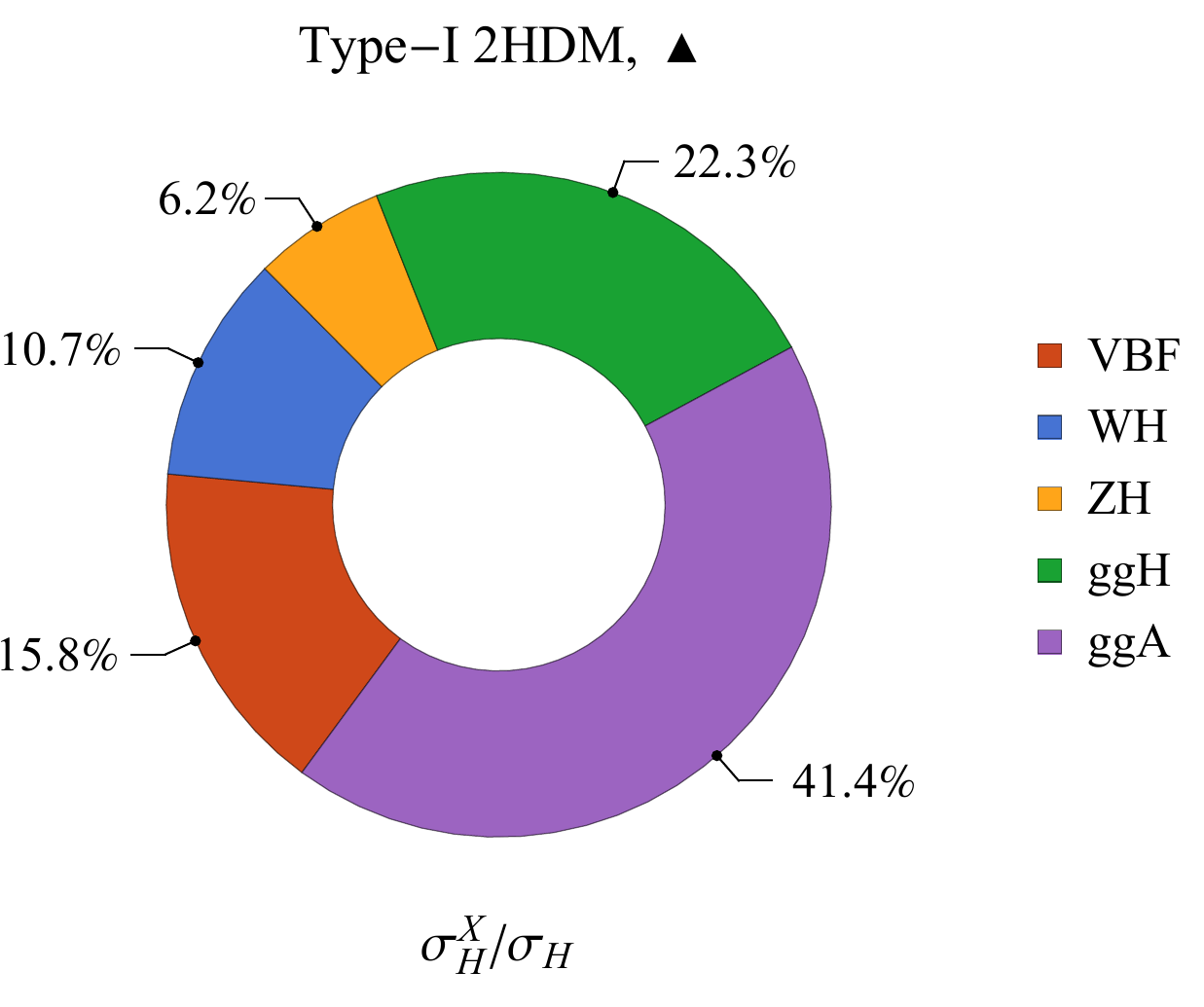}  \qquad  
\includegraphics[width=0.45\textwidth]{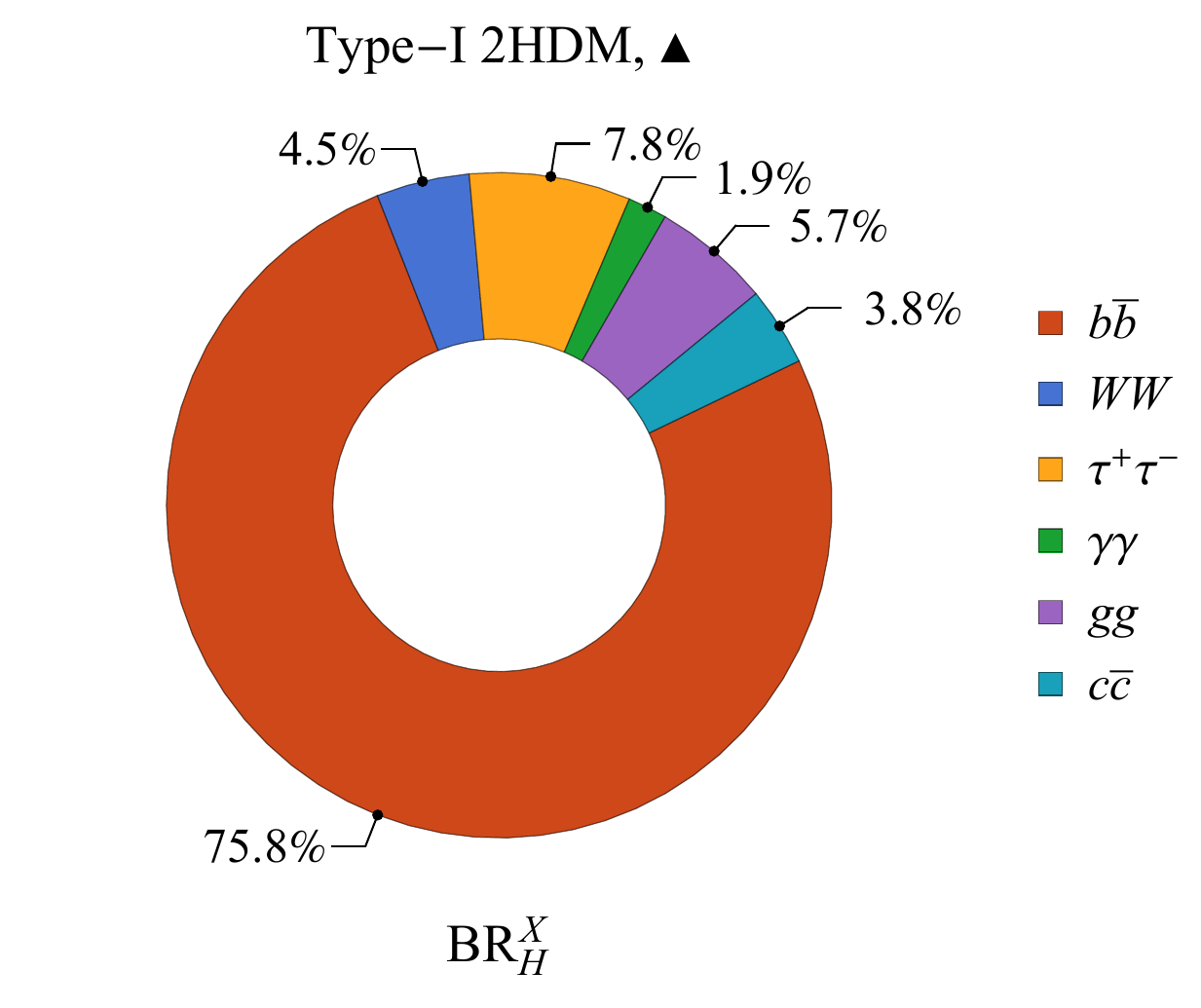}  
\vspace{2mm}
\caption{\label{fig:10} As Figure~\ref{fig:2} but for the third type-I~2HDM   benchmark scenario~(\ref{eq:bench3typeI}). The label ``ggA''  in the left panel refers to $H$ production through $gg \to A$ followed by $A \to W^{\mp} H^{\pm} \to W^{\mp} W^{\pm} H$.  Apart from ${\rm BR}_H^{\gamma \gamma}$ only values of $\sigma_H^X/\sigma_H$ and ${\rm BR}_H^X$ that exceed 2\% are shown.} 
\end{center}
\end{figure}

The percentage breakdown of  the different $H$ production channels and the values of the branching ratios of $H$ for the type-I~2HDM parameter scenario~(\ref{eq:bench3typeI}) are presented in Figure~\ref{fig:10}.  The left pie chart shows that $41.4\%$ of $\sigma_H = 3.6 \, {\rm pb}$  are due to $A$ production in gluon-fusion (ggA) with $A \to W^{\mp} H^{\pm} \to W^{\mp} W^{\pm} H$. An example of a Feynman diagram that gives rise to this exotic $H$ production mode is displayed in the middle of Figure~\ref{fig:1}. The combination of the VBF, WH and ZH channels (the ggH channel itself) is instead subleading and amounts to $32.7\%$~($22.3\%$). 

\begin{figure}[!t]
\begin{center}
\includegraphics[width=0.45\textwidth]{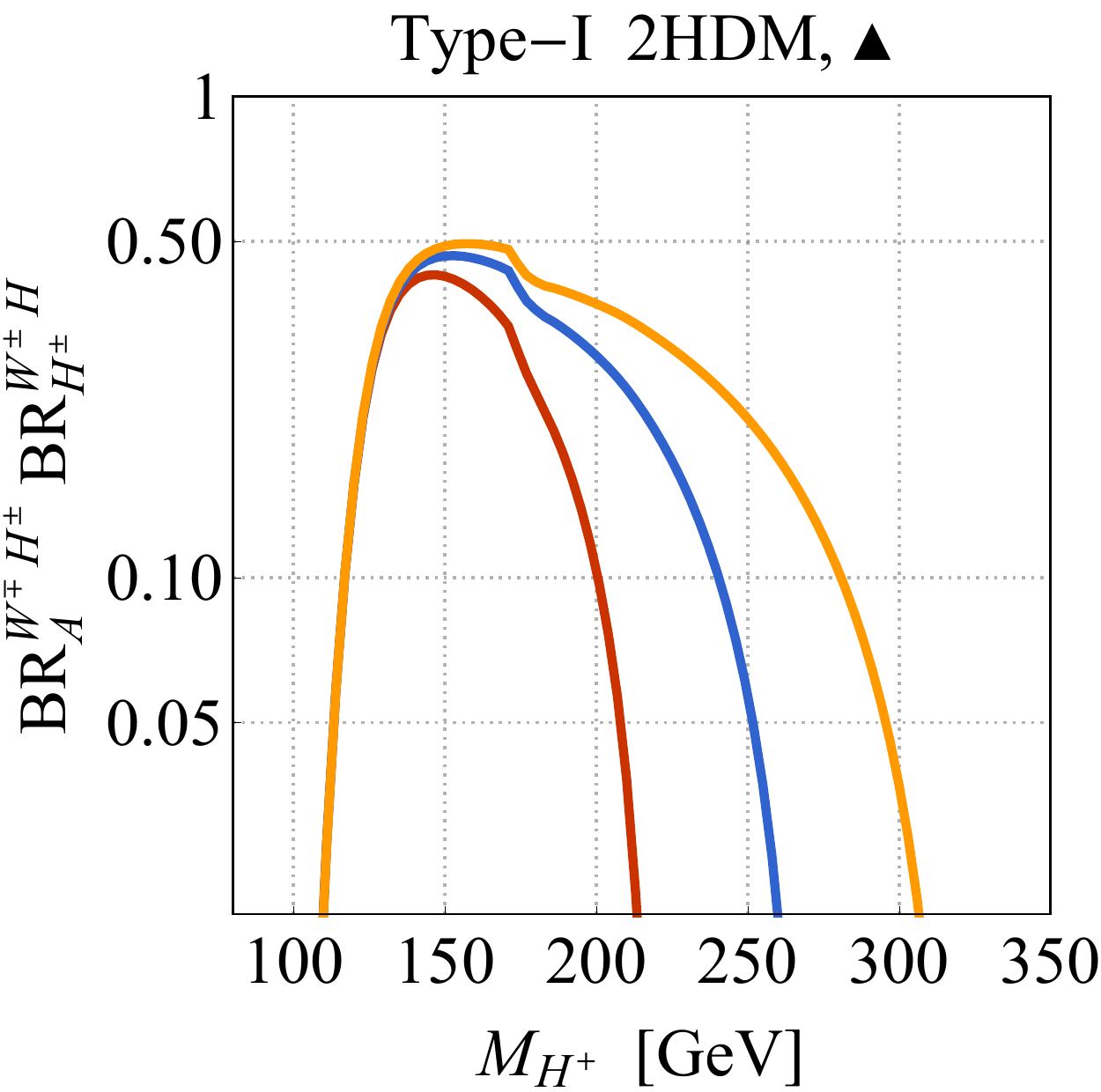}  \qquad  
\includegraphics[width=0.45\textwidth]{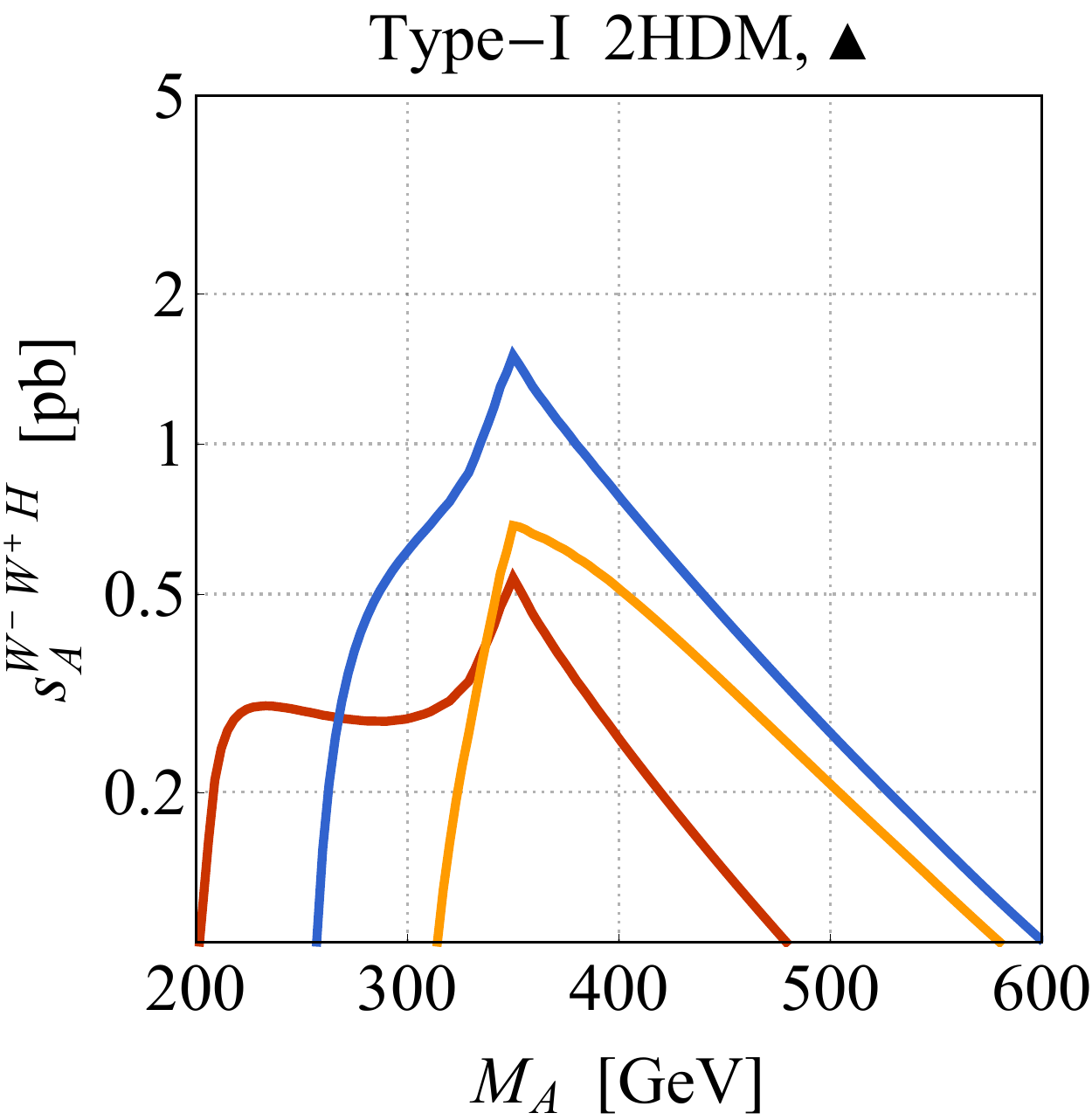}  
\vspace{2mm}
\caption{\label{fig:11} Left: The product of the branching ratios $A \to W^{\mp} H^{\pm}$ and $H^{\pm} \to W^{\pm} H$ as a function of $M_{H^+}$. The red, blue and yellow curve correspond to $M_A = 300 \, {\rm GeV}$, $350 \, {\rm GeV}$ and $400 \, {\rm GeV}$. Right:  Signal strength at $\sqrt{s} = 13 \, {\rm TeV}$ for $pp \to A \to W^- W^+ H$  as a function of $M_A$. The red, blue and yellow curve correspond to $M_{H^+} = 120 \, {\rm GeV}$, $170 \, {\rm GeV}$ and $220 \, {\rm GeV}$. All results have been obtained in the benchmark scenario~(\ref{eq:bench3typeI}).}
\end{center}
\end{figure}

In Figure~\ref{fig:11} we show results for ${\rm BR}_A^{W^{\mp} H^{\pm}} \, {\rm BR}_{H^{\pm}}^{W^{\pm}H}$ and $s_{A}^{W^-W^+ H}$ as a function of $M_{H^+}$ and $M_A$, respectively. From the left panel one observes that the product ${\rm BR}_A^{W^{\mp} H^{\pm}} \, {\rm BR}_{H^{\pm}}^{W^{\pm}H}$  of branching ratios exceeds 10\% for $M_{H^+} \underset{\text{\relsize{+2}$\tilde{}$}}{\in} [120, 280] \, {\rm GeV}$, reaching a peak value of almost 50\% at around $160 \, {\rm GeV}$. It follows that the signal strength $s_A^{W^- W^+ H}$  in $pp \to A \to W^- W^+ H$ production at $\sqrt{s} = 13 \, {\rm TeV}$ can reach the level of $1 \, {\rm pb}$ for $M_A \simeq 2 M_{H^+} \simeq 350 \, {\rm GeV}$. This feature is illustrated by the plot on the right-hand side in Figure~\ref{fig:11}. The branching ratios of $H$ in the benchmark scenario~(\ref{eq:bench3typeI}) are given in the right panel of Figure~\ref{fig:10}. The two largest branching ratios of $75.8\%$ and $7.8\%$ are those to bottom  and tau pairs, while the di-photon branching ratio amounts to only $1.9\%$. The corresponding signal strengths are  $s_H^{b \bar b} = 2.8 \, {\rm pb}$,  $s_H^{\tau^+ \tau^-} = 0.28 \, {\rm pb}$ and  $s_H^{\gamma \gamma} = 0.07 \, {\rm pb}$ at $\sqrt{s} = 13 \, {\rm TeV}$.

It remains to be verified that the $H$, $A$ and $H^+$ featured in our third type-I~2HDM parameter scenario are phenomenologically viable.  In this context, we first note  that the sensitivity of the ATLAS  di-photon search at $\sqrt{s} = 8 \, {\rm TeV}$~\cite{Aad:2014ioa} is by a factor of approximately $1.8$ too low to probe the parameter choices~(\ref{eq:bench3typeI}). Likewise,  di-tau searches such as~\cite{CMS:2016rjp}  provide no relevant constraints.  In the case of the $A$, one observes from the lower left panel in Figure~\ref{fig:4a} that for $M_A = 350 \, {\rm GeV}$ the ATLAS search for $A \to Zh$~\cite{Aaboud:2017cxo}  requires $\tan \beta > 3.3$. The 95\%~CL bound on $\tan \beta$ that follows from the $B_s \to \mu^+ \mu^-$ measurements of CMS and LHCb~\cite{CMS:2014xfa,Aaij:2017vad} reads $\tan \beta > 3.3$  for $M_{H^+}  = 170 \, {\rm GeV}$ --- see the left panel in  Figure~\ref{fig:4b}. At present the $M_A$, $M_{H^+}$ and $\tan \beta$ values chosen in~(\ref{eq:bench3typeI}) are thus allowed. Future~LHC searches for $A \to Zh/H$ and/or $B_s \to \mu^+ \mu^-$ should however be able to probe model realisations that feature  parameters not much different from~(\ref{eq:bench3typeI}). 

The predictions shown in Figure~\ref{fig:9} and in the right pie chart  of Figure~\ref{fig:10} again include the contributions of charged Higgs loops to $\Gamma \left (h \to \gamma \gamma \right )$ and $\Gamma \left (H \to \gamma \gamma \right )$.  We find that charged Higgs effects suppress the  di-photon $h$ and $H$ decay rates by $10\%$ and $20\%$ compared to the case with only top-quark  and $W$-boson contributions. Numerically, we obtain $\kappa_\gamma^h = 0.78$. To conclude the discussion of the third benchmark scenario, we mention that for the choice of parameters employed in~(\ref{eq:bench3typeI}) the Higgs potential  is bounded from below and the constraint $\Delta \rho \in [-1.2, 2.4] \cdot 10^{-3}$ that follows  from the electroweak precision measurements is satisfied. 

\subsection{Square benchmark scenario}
\label{sec:bench4typeI}

The parameter choices in our fourth and final type-I~2HDM benchmark scenario are 
\begin{equation} \label{eq:bench4typeI} 
\sin \alpha  = 0.05  \,, \hspace{1mm}
\tan \beta = 4.2 \,,  \hspace{1mm}
M_H = 95 \, {\rm GeV} \,, \hspace{1mm}
M_A = 80 \, {\rm GeV} \,, \hspace{1mm}
M_{H^+} = 87 \, {\rm GeV}   \,, \hspace{1mm}
\lambda_3 = 0.26 \,. 
\end{equation}

\begin{figure}[!t]
\begin{center}
\includegraphics[width=0.45\textwidth]{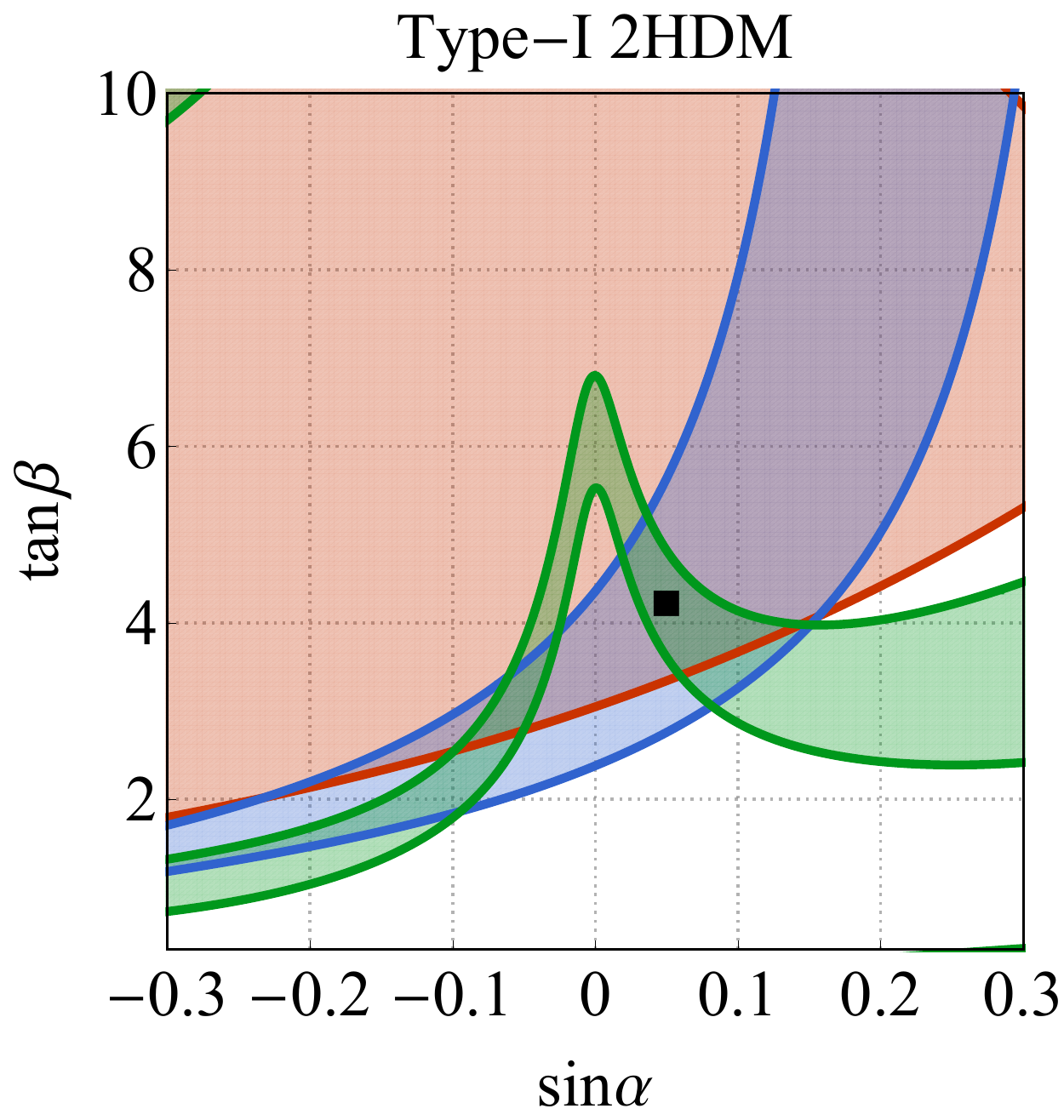}  
\vspace{2mm}
\caption{\label{fig:12}  Allowed/favoured parameter regions in the case of the square ($\blacksquare$) type-I~2HDM benchmark scenario~(\ref{eq:bench4typeI}). The colour coding resembles that of Figure~\ref{fig:2}.}
\end{center}
\end{figure}

The allowed parameter regions corresponding to~(\ref{eq:bench4typeI}) are displayed in Figure~\ref{fig:12}. The red, blue and green contours enclose the parameters that a preferred by the LHC~Run-I Higgs data, the LEP excess in $e^+ e^- \to ZH$ and the CMS di-photon anomaly, respectively.  As before the parameters $M_H$, $M_A$, $M_{H^+}$ and $\lambda_3$ have been kept fixed when calculating the constraints. The values of  $\sin \alpha$ and $\tan \beta$ as chosen in~(\ref{eq:bench4typeI}) are indicated by a square, and one observes that  these parameters lead to a consistent overall picture. 

\begin{figure}[!t]
\begin{center}
\includegraphics[width=0.45\textwidth]{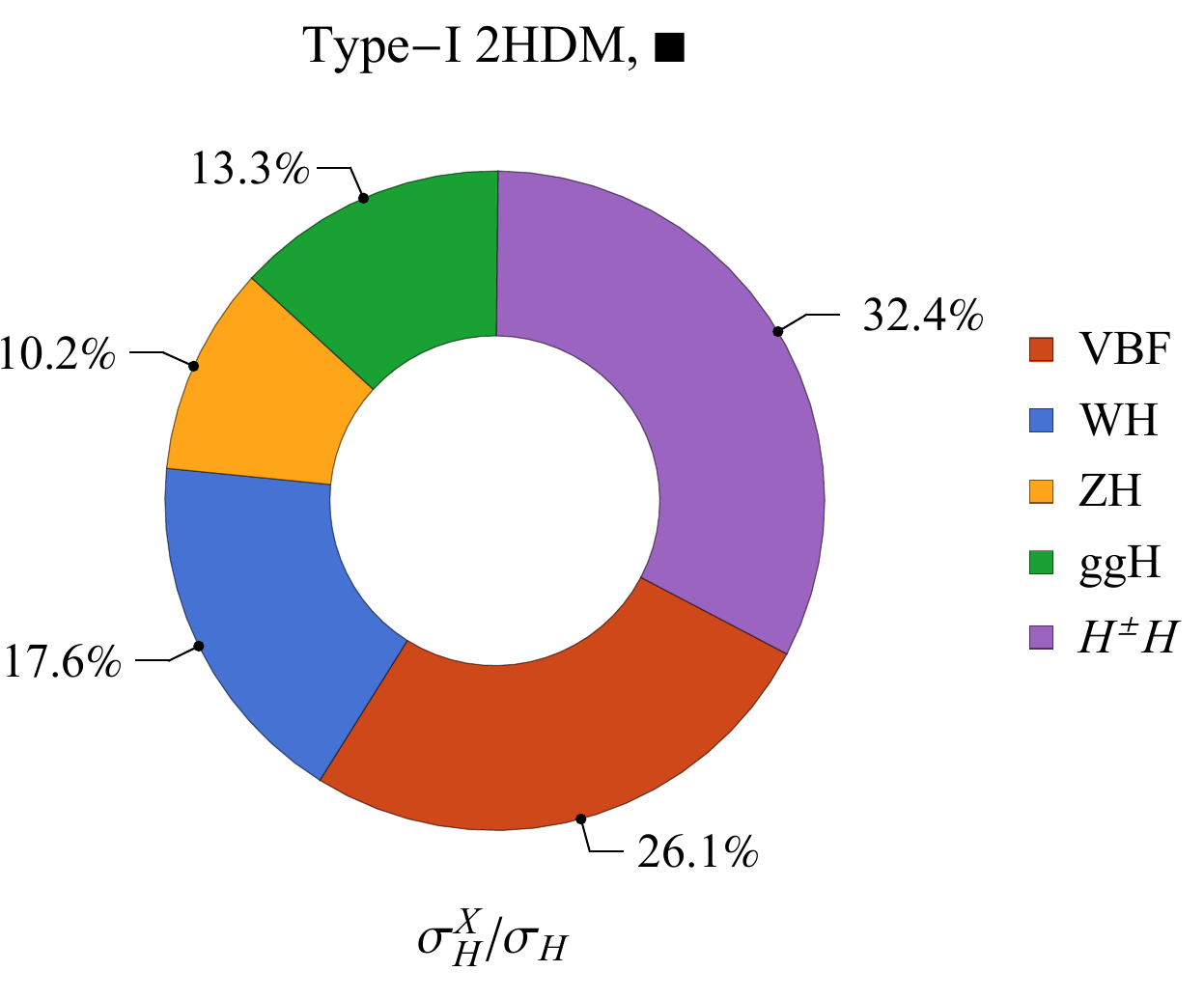}  \qquad  
\includegraphics[width=0.45\textwidth]{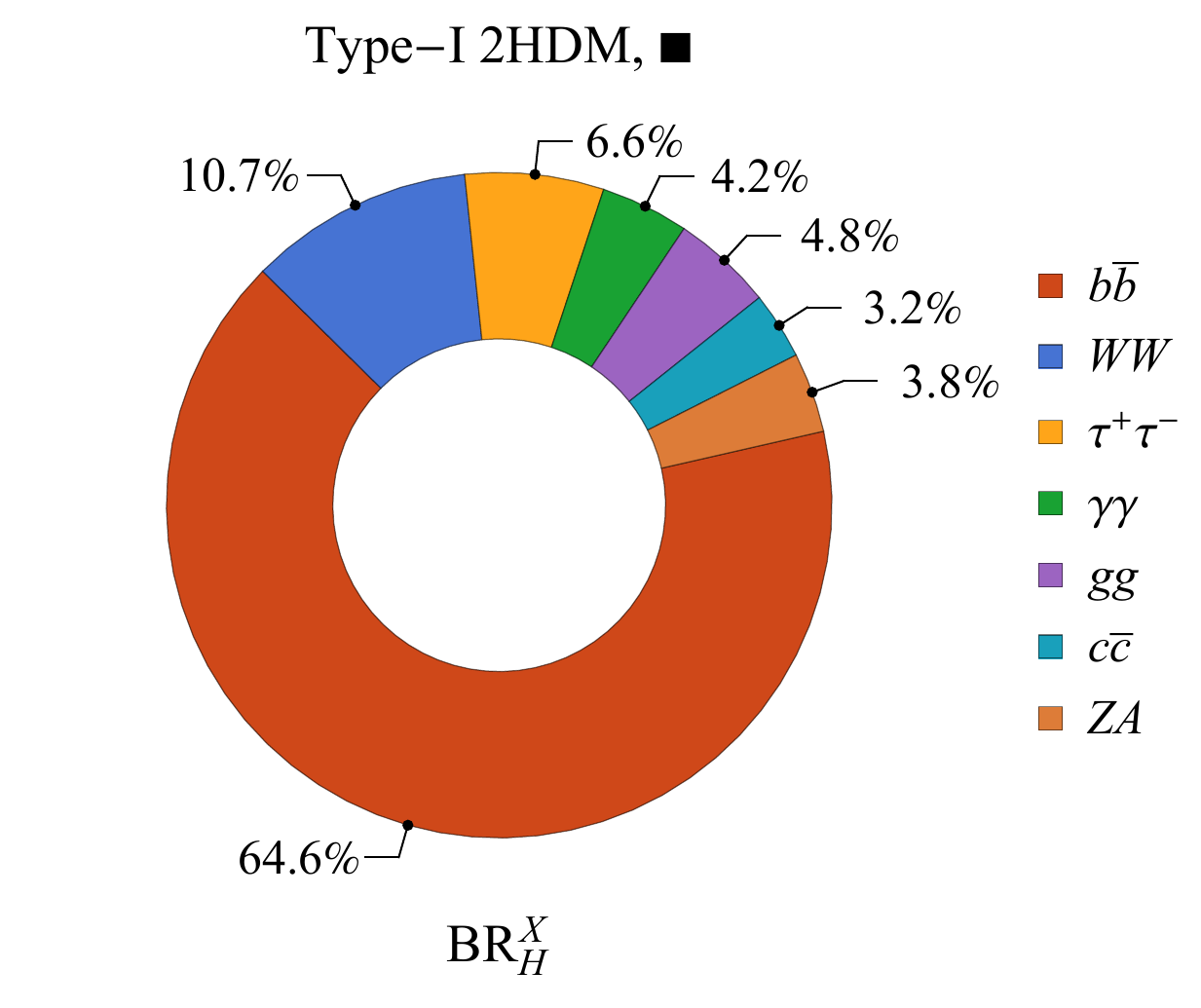}  
\vspace{2mm}
\caption{\label{fig:13} As Figure~\ref{fig:2} but for the fourth type-I~2HDM   benchmark scenario~(\ref{eq:bench4typeI}). The label ``$H^\pm H$''  in the left panel refers to associated $H$ production via $pp \to W^{\pm \, \ast} \to H^{\pm} H$. Only values of  ${\rm BR}_H^X$ that are larger than 2\% are explicitly given. }
\end{center}
\end{figure}

In Figure~\ref{fig:13} we present the breakdown of  the different $H$ production channels and the values of the branching ratios of $H$ for the fourth parameter scenario~(\ref{eq:bench4typeI}). An inspection of the left pie chart reveals that $32.4\%$ of $\sigma_H = 1.5 \, {\rm pb}$  stem from associated $H^\pm H$ production. A graph that contributes to this production mode  is shown on the right-hand side in Figure~\ref{fig:1}.  We calculate the relevant cross section with {\tt MadGraph5\_aMC\@NLO}~\cite{Alwall:2014hca}  at next-to-leading order in QCD using an {\tt UFO} implementation~\cite{Degrande:2011ua} of the 2HDM model discussed in~\cite{Bauer:2017ota}.  It follows that  for the parameter choices~(\ref{eq:bench4typeI}), $H$ production through $pp \to W^{\pm \, \ast} \to H^\pm H$ is almost as important as the combination of the VBF, WH and ZH channels which gives rise to $53.9\%$ of the total cross section. As before $H$ production via ggH  is only of very limited importance. 

The dominance of $H^\pm H$ production is readily understood by noticing that the ratio between the $W^\pm H^\mp H$ and $W^\pm W^\mp H$ or $Z ZH$ coupling is simply given by $\tan^2 \left ( \beta - \alpha \right )$. For a sufficiently fermiophobic $H$ and very light $H$ and $H^+$ states, the $H$ production rate in $pp \to W^{\pm \, \ast} \to H^\pm H$ can thus be comparable to or even larger than the VBF, WH and ZH modes taken together~\cite{Akeroyd:2003bt,Akeroyd:2003xi}. Our results for the $H$ branching ratios corresponding to~(\ref{eq:bench4typeI}) are displayed on the right-hand side in Figure~\ref{fig:13}. The six largest branching ratios are those to bottom quarks~($64.6\%$), $W$~bosons~($10.7\%$), $\tau^+ \tau^-$ pairs~($6.6\%$), gluons~($4.8\%$), photons~($4.2\%$) and $ZA$~pairs~($3.8\%$). These predictions lead to  signal strengths of  $s_H^{b \bar b} = 0.98 \, {\rm pb}$,  $s_H^{WW} = 0.16 \, {\rm pb}$,    $s_H^{\tau^+ \tau^-} = 0.10 \, {\rm pb}$, $s_H^{gg} = 0.07 \, {\rm pb}$, $s_H^{\gamma \gamma} = 0.06 \, {\rm pb}$ and $s_H^{ZA} = 0.06 \, {\rm pb}$ at $\sqrt{s} = 13 \, {\rm TeV}$. 

Like in the case of the other three benchmark scenarios discussed earlier, it is easy to show that apart from~\cite{CMS-PAS-HIG-17-013}  there do not exist direct measurements that are sensitive to a~$H$ with properties similar to~(\ref{eq:bench4typeI}). The only existing LHC analyses that allow to constrain a very light~$A$ with a mass of $80 \, {\rm GeV}$ are the $A \to \gamma \gamma$ searches~\cite{CMS-PAS-HIG-17-013,Aad:2014ioa}. The strongest limit on $\tan \beta$ as a function of $M_A$ follows from the $\sqrt{s} = 13 \, {\rm TeV}$ results of CMS that are based on an integrated luminosity of $35.9 \, {\rm fb}^{-1}$. The corresponding constraint is indicated in  the lower right panel of Figure~\ref{fig:4a} by the red curve. One sees that for $M_A = 80 \, {\rm GeV}$  only values of $\tan \beta < 0.4$ are excluded at 95\%~CL. The benchmark scenario~(\ref{eq:bench4typeI}) however employs $\tan \beta = 4.2$ and is thus clearly allowed. We also mention that  $\Gamma \left (h \to Z^\ast A \right )$ is far too small to be subject to the existing indirect LHC constraints on the total Higgs decay width~$\Gamma_h$~(the relevant $\sqrt{s} = 13 \, {\rm TeV}$ results can be found in~\cite{Aad:2015xua,Khachatryan:2016ctc}).

One finally needs to check that a $H^+$  with a mass of $87 \, {\rm GeV}$ is consistent with all direct and indirect constraints in the type-I~2HDM. Strong lower bounds on $M_{H^+}$ arise from LEP searches for pair-produced charged Higgs bosons~\cite{Abbiendi:2013hk}. We find that in our type-I~2HDM benchmark scenario~(\ref{eq:bench4typeI}) charged Higgs masses below $86.9 \, {\rm GeV}$ are excluded at 95\%~CL. At low mass direct limits also arise from LHC searches for $H^+ \to \tau^+ \nu_\tau$ with the latest results given in~\cite{ATLAS-CONF-2016-088, CMS-PAS-HIG-16-031}. As can be seen from the panels in Figure~\ref{fig:4b}, in the range $M_{H^+} \underset{\text{\relsize{+2}$\tilde{}$}}{\in} [85, 105] \, {\rm GeV}$ the $H^+ \to \tau^+ \nu_\tau$ search~\cite{CMS-PAS-HIG-16-031} in fact provides presently the strongest constraint on $\tan \beta$. Numerically, we find for $M_{H^+} = 87 \, {\rm GeV}$ the bound $\tan \beta > 4.1$, which does not rule out the choice of $\tan \beta$ made in~(\ref{eq:bench4typeI}). Improved LHC searches for $H^+ \to \tau^+ \nu_\tau$ should however be able to exclude or find evidence for scenarios with $\tan \beta \simeq 5$ and a charged Higgs of mass close to $M_Z$.

One motivation for a $H^+$  with a mass close to $M_W$ is that such a state  can partly explain the $2.3\sigma$ excess~\cite{Patrignani:2016xqp} observed in the  lepton-flavour universality ratio $R_{\tau/\ell} = 2 \hspace{0.25mm} {\rm BR}_{W+}^{\tau^+ \nu_\tau}/\sum_{\ell=e,\mu} {\rm BR}_{W+}^{\ell^+ \nu_\ell}$. Using the results of~\cite{Park:2006gk}, we in fact find  that  the deviation in $R_{\tau/\ell}$ is reduced to $1.9\sigma$ in the  benchmark scenario~(\ref{eq:bench1typeI}) as a result of the contamination of the $W^+ \to \tau^+ \nu_\tau$ signal by $H^+ \to \tau^+ \nu_\tau$ decays. We furthermore note that while most LEP searches focus on the $H^+ \to \tau^+\nu_\tau$ and $H^+ \to c \bar s$ channels also $H^+ \to W^{+} A \,  (A \to b \bar b)$ has been considered to  cover the possibility of pseudoscalars with $M_A < 70 \, {\rm GeV}$. For such scenarios the $M_{H^+}$-limits weaken and we observe  that in the type-I~2HDM only charged Higgs masses below $81.4 \, {\rm GeV}$ are excluded at 95\%~CL if $M_A$ is taken to be $50 \, {\rm GeV}$. In such a case the deviation in $R_{\tau/\ell}$ would be reduced to $1.4\sigma$. Although a $A$ with $50 \, {\rm GeV}$ can be shown to pass the direct constraints from $h \to AA \to \mu^+ \mu^- b \bar b$~\cite{Khachatryan:2017mnf} as well as the indirect bounds from the LHC searches for off-shell $h$ production~(see~\cite{Aad:2015xua,Khachatryan:2016ctc} for example),  we do not consider the case $M_A = 50 \, {\rm GeV}$ here. The reason is that such a choice does not allow to simultaneously explain   the LEP anomaly in $e^+ e^- \to ZH$ and the CMS excess in $H \to \gamma \gamma$, as a result of the large partial $H \to Z^{\ast} A$ decay width that suppresses ${\rm BR}_H^{\gamma \gamma}$. We finally add that a precision measurement of  $R_{\tau/\ell}$ is challenging for ATLAS and CMS due to triggering and the uncertain tau identification efficiency, but may be possible at LHCb by performing a dedicated analysis~\cite{Albrecht:2013fba,LHCbVELOGroup:2014uea}. 

As in all other 2HDM benchmark scenarios the results displayed in Figure~\ref{fig:12} and in the right pie chart of Figure~\ref{fig:13} take into account charged Higgs contributions to $\Gamma \left (h \to \gamma \gamma \right )$ and $\Gamma \left (H \to \gamma \gamma \right )$. These corrections lead to a suppression of around $10\%$ ($30\%$) for a $h$ ($H$)  relative to the case with only top-quark  and $W$-boson contributions, resulting in $\kappa_\gamma^h = 0.80$. We have also verified that for the parameters employed in~(\ref{eq:bench4typeI}) the Higgs potential  is bounded from below and that the constraints that arise from the $\rho$~parameter are fulfilled at $2\sigma$. 

\section{Conclusions}
\label{sec:conclusions}

In 2016 the ATLAS and CMS collaborations each have collected around $40 \, {\rm fb}^{-1}$ of LHC data at  $\sqrt{s} = 13 \, {\rm TeV}$. While most of the measurements they have performed are in full agreement with the corresponding SM predictions some glitches have been observed.  For instance there  are  excesses in the multi-lepton channel of $t \bar t h$ production at about  $2\sigma$~\cite{ATLAS-CONF-2016-058,CMS-PAS-HIG-16-022} and an unexpected bump at around $95 \, {\rm GeV}$ in the di-photon mass spectrum~\cite{CMS-PAS-HIG-17-013,CMS-PAS-HIG-14-037} with a global (local) significance of $2.8\sigma$~($1.3\sigma$). Although none of these deviations is on its own statistically significant, it seems like an interesting and useful exercise to try to understand if these anomalies can arise in a coherent way from physics beyond the SM. 

In our article, we have shown that the type-I~2HDM can provide a very economic explanation of both the multi-lepton and di-photon excess observed at LHC,  while simultaneously addressing two historic $2\sigma$ Higgs  anomalies that linger around since the times of LEP~\cite{Barate:2003sz} and the Tevatron~\cite{Collaboration:2012bk,Aaltonen:2013ipa}. The key ingredient to describe the observed Higgs excesses is a moderately-to-strongly fermiophobic CP-even Higgs $H$ with a mass of $95 \, {\rm GeV}$. Due to its fermiophobic nature such a $H$ has an enhanced di-photon branching ratio making it possible to obtain a signal strength  of the order of~$0.1 \, {\rm pb}$ by the combination of VBF, WH and ZH production alone. A sizeable $H$ production rate can however also arise from either  top-quark pair and single-top production followed by $\overset{\text{\relsize{-2}(}-\text{\relsize{-2})}}{t} \to H^{\pm} \overset{\text{\relsize{-2}(}-\text{\relsize{-2})}}{b} \to W^{\pm \, \ast} H \overset{\text{\relsize{-2}(}-\text{\relsize{-2})}}{b}$ or from ggA production with  $A \to W^{\mp} H^{\pm} \to W^{\mp} W^{\pm} H$. In cases where the~$H$ is strongly fermiophobic and  the charged Higgs is very light the process $pp \to W^{\pm \, \ast} \to H^\pm H$  can finally provide an efficient way to produce the non-SM CP-even Higgs. 

By means of a detailed  numerical analysis we have then demonstrated   that all the considered Higgs excesses can be simultaneously reproduced if $H$ production is dominated by $t \bar t$~production followed by the cascade $\overset{\text{\relsize{-2}(}-\text{\relsize{-2})}}{t} \to H^{\pm} \overset{\text{\relsize{-2}(}-\text{\relsize{-2})}}{b} \to W^{\pm \, \ast} H \overset{\text{\relsize{-2}(}-\text{\relsize{-2})}}{b}$. This option can be realised in the type-I~2HDM in parameter regions with $M_{H^+} \simeq 130 \, {\rm GeV}$ and~$\tan \beta \in [4,6]$. If the inclusive $H$  cross section instead  receives the largest contribution from VBF production, ggA~production  with  $A \to W^{\mp} H^{\pm} \to W^{\mp} W^{\pm} H$ or associated $H^\pm H$ production only the CMS excess in $H \to \gamma \gamma$  and the LEP anomaly in $e^+ e^- \to ZH$ can be explained, while the deviations seen in the $t \bar t h$ channel remain unaccounted for. We find that in order for $pp \to W^{\pm \, \ast} \to H^\pm H$ to be the leading production mechanism one has to have $M_{H^+} \lesssim 100 \, {\rm GeV}$, whereas  cascade $H$ production initiated by $gg \to A$ is the dominant production channel for $M_{H^+} \simeq M_A/2 \simeq 170 \, {\rm GeV}$.  The sum of the VBF, WH and ZH channels can finally give a sizeable  inclusive $H$ cross section  even for moderately heavy  charged Higgses. We~stress that one firm conclusion that can be drawn from our analysis is that in the considered new-physics model any di-photon excess should be associated with additional detector activity such as forward or bottom-quark jets. This feature should provide useful handles in future LHC analyses to  improve the separation of new-physics signal and SM backgrounds.

All type-I~2HDM realisations that we have explored in our work include other light Higgses besides $H$. The current constraints from direct and indirect searches for spin-0 resonances can however be shown to be satisfied for the four benchmark scenarios that we have discussed in detail. Future LHC searches for charged Higgses in the $H^+ \to \tau^+ \nu_\tau$ channel  or improved measurements of flavour observables such as $B_s \to \mu^+ \mu^-$ should nevertheless be able to exclude parts of the parameter space that leads to a simultaneous explanation of the discussed anomalies. This statement is particularly true for model realisations that lead to sizeable $H^\pm H$ production rates or exotic Higgs signatures involving the decay chains $\overset{\text{\relsize{-2}(}-\text{\relsize{-2})}}{t} \to H^{\pm} \overset{\text{\relsize{-2}(}-\text{\relsize{-2})}}{b} \to W^{\pm \, \ast} H \overset{\text{\relsize{-2}(}-\text{\relsize{-2})}}{b}$ or  $A \to W^{\mp} H^{\pm} \to W^{\mp} W^{\pm} H$, since~in all these cases the charged Higgs has to be necessarily light. The best search strategy for the $A$ depends strongly on its mass. For pseudoscalars with $M_A \lesssim 160 \, {\rm GeV}$, we find that the channels $A \to \gamma \gamma$ and $A \to \tau^+ \tau^-$ offer only limited sensitivity to $\tan \beta$ values significantly above  1. Better prospects to probe the fermiophobic type-I 2HDM scenarios discussed in our article seem to be provided by $A \to Zh/H$ searches, which already now furnish the leading restrictions on~$\tan \beta$ for larger pseudoscalar masses.  

\acknowledgments
We thank Andreas~Crivellin for enlightening conversations during the preparation of this article, including discussions concerning his recent paper~\cite{Crivellin:2017upt}, and for useful comments on the manuscript. We are also grateful to  Martin~Bauer for interesting discussions concerning electroweak precision measurements and his encouragement, and would like to thank Fady Bishara for help with~{\tt ROOT} as well as  Chris~Hays and Mika~Vesterinen for useful communications concerning lepton non-universality in $W$ decays. We are  thankful as well  to Andrew~Akeroyd, Giacomo~Cacciapaglia, Junjie Cao and Felix Kling for their positive feedback and making us aware of~\cite{Cacciapaglia:2016tlr, Akeroyd:2003bt,Akeroyd:2003xi,Akeroyd:1998dt,Coleppa:2014hxa,Coleppa:2014cca,Kling:2015uba,Arhrib:2016wpw,Cao:2016uwt}. We~finally would like to express our gratitude to Andrew~Akeroyd for drawing our attention to the process $pp \to W^{\pm \, \ast} \to H^\pm H$ and pointing out that under certain circumstances it can be phenomenologically relevant.  UH~appreciates the continued hospitality and support of the CERN Theoretical Physics Department.  

%\providecommand{\eprint}[2][]{\href{https://arxiv.org/abs/#2}{\texttt{#2}}}

%\bibliographystyle{apsrev}
%\bibliography{paper}

\end{document}